%% file: main.tex
\documentclass[acmlarge, nonacm]{acmart}
\renewcommand\footnotetextcopyrightpermission[1]{} % 
\usepackage{soul}
\usepackage{tabularx}
\usepackage{array,multirow,graphicx}
\usepackage{threeparttable}
\usepackage{tabulary}

\AtBeginDocument{
  \providecommand\BibTeX{{
    \normalfont B\kern-0.5em{\scshape i\kern-0.25em b}\kern-0.8em\TeX}}}

\setcopyright{none}
\begin{document}

\title{Towards Automating Retinoscopy for Refractive Error Diagnosis}

\author{Aditya Aggarwal}
\orcid{0000-0001-8862-2581}
\email{aditya097.aggarwal@gmail.com}

\author{Siddhartha Gairola}
\orcid{0000-0001-7098-1703}

\author{Uddeshya Upadhyay}
\orcid{0000-0002-9783-7063}
\affiliation{
    \institution{Microsoft Research}
    \country{India}
}

\author{Akshay P Vasishta}
\orcid{0000-0001-5988-163X}

\author{Diwakar Rao}
\orcid{0000-0001-7538-1184}

\author{Aditya Goyal}
\orcid{0000-0003-1126-2594}

\author{Kaushik Murali}
\orcid{0000-0002-1385-3227}
\affiliation{
   \institution{Sankara Eye Hospital}
   \country{India}
 }
 
\author{Nipun Kwatra}
\orcid{0000-0003-0354-6204}
\email{nkwatra@microsoft.com}

\author{Mohit Jain}
\orcid{0000-0002-7106-164X}
\email{mohja@microsoft.com}
\affiliation{
  \institution{Microsoft Research}
  \country{India}
}

\renewcommand{\shortauthors}{Aggarwal et al.}

\input{sections/0_abstract}

\maketitle

\input{sections/1_introduction}
\input{sections/2_background}
\input{sections/3_relatedWork}
\input{sections/4_dataCollection}

\input{sections/5_retinoscopyDerivation}
\input{sections/6_imageProcessing}

\input{sections/7_results}
\input{sections/8_discussion}
\input{sections/9_conclusion}

\begin{acks}
We would like to thank optometrists Amaravathi P. and Wartson for helping with the data collection, Banukumar Rajendran for helping with retinoscope-smartphone assembly, and all the participants for their time and patience.
\end{acks}

\bibliographystyle{format/ACM-Reference-Format}
\bibliography{references}

\appendix

\end{document}

%% file: sections/0_abstract.tex
\begin{abstract}
Refractive error is the most common eye disorder and is the key cause behind correctable visual impairment, responsible for nearly 80\% of the visual impairment in the US.
Refractive error can be diagnosed using multiple methods, including subjective refraction, retinoscopy, and autorefractors.
Although subjective refraction is the gold standard, it requires cooperation from the patient and hence is not suitable for infants, young children, and developmentally delayed adults. 
Retinoscopy is an objective refraction method that does not require any input from the patient. However, retinoscopy requires a lens kit and a trained examiner, which limits its use for mass screening.
In this work, we automate retinoscopy by attaching a smartphone to a retinoscope and recording retinoscopic videos with the patient wearing a custom pair of paper frames.
We develop a video processing pipeline that takes retinoscopic videos as input and estimates the net refractive error based on our proposed extension of the retinoscopy mathematical model. Our system alleviates the need for a lens kit and can be performed by an untrained examiner.
In a clinical trial with 185 eyes, we achieved a sensitivity of 91.0\% and specificity of 74.0\% on refractive error diagnosis. Moreover, the mean absolute error of our approach was 0.75$\pm$0.67D on net refractive error estimation compared to subjective refraction measurements.
Our results indicate that our approach has the potential to be used as a retinoscopy-based refractive error screening tool in real-world medical settings.
\end{abstract}


\ccsdesc[300]{Human-centered computing~Ubiquitous and mobile computing}
\ccsdesc[500]{Applied computing~Consumer health}
\keywords{Retinoscopy, retinoscope, myopia, hyperopia, health sensing, refractive error, screening, diagnosis, smartphone, image processing, mobile health, low cost system, optics}

%% file: sections/1_introduction.tex
\section{Introduction} \label{sec:intro}

Refractive error is the most prevalent eye disorder~\cite{WHO_vision_report_2019}, where the eye is unable to bend light correctly to focus on the retina, resulting in blurred vision. There are four types of refractive error~\cite{Optham_book}--myopia, hyperopia, astigmatism, and presbyopia. There are three reasons which can lead to them. First, an eyeball that is too long or too short can make it hard for the eye to focus, leading to myopia (nearsightedness) and hyperopia (farsightedness), respectively. 
Second, a distortion in the shape of the cornea, which contributes to ${\sim}70\%$ of the eye's refractive power, can lead to astigmatism. 
Third, the eye lens gradually thickens with age and looses its flexibility, making it hard to focus, leading to presbyopia. Refractive error is not a preventable disorder~\cite{WHO_vision_report_2019}; however, it can be treated using corrective glasses or contact lenses if diagnosed early~\cite{Optham_book}.

Uncorrected refractive error is one of the leading causes of vision impairment~\cite{Pascolini614}. In 2010, uncorrected refractive errors were found to be responsible for visual impairment in 101.2 million people and blindness\footnote{WHO
defines blindness as visual acuity worse than 3/60, \textit{i.e.}, a person with blindness needs to be as close as 3 feet to see what a person with normal vision can see at 60 feet~\cite{WHO_vision_report_2019}.
} in 6.8 million people~\cite{Naidoo2016-hc}. Blindness due to undiagnosed refractive error is more prevalent in the global south, \textit{e.g.}, in India, it is 1.06\% among people aged >40 years~\cite{india-referror}, compared to 0.24\% in the USA~\cite{us-referror} and 0.26\% in the UK~\cite{uk-reference}.
Undiagnosed refractive error can also cause other severe eye disorders, such as amblyopia in children, in which high refractive error in one eye causes the brain to ignore the blurred image from the weaker eye. This can lead to permanent vision loss of the weaker eye if not corrected in time. 
Vision screening programs at a large community scale are thus important to address this preventable source of blindness. However, screenings are difficult in developing countries due to a combination of factors, such as poor access to eye care services, unavailability of affordable solutions, and lack of awareness. The conditions are further exacerbated in low-income regions like Central Africa, where less than six trained eye specialists\footnote{Eye specialists comprise of ophthalmologists, optometrists, and allied ophthalmic personnel.} are available for a million people~\cite{30_years_trends}. Therefore, tools to enable refractive error screening in a low-cost and portable manner, without the need of skilled practitioners, can help in democratizing such mass screening programs. 
Our work is a step in that direction.

\begin{figure*}
\begin{center}
    \centering
    \includegraphics[width=1\linewidth]{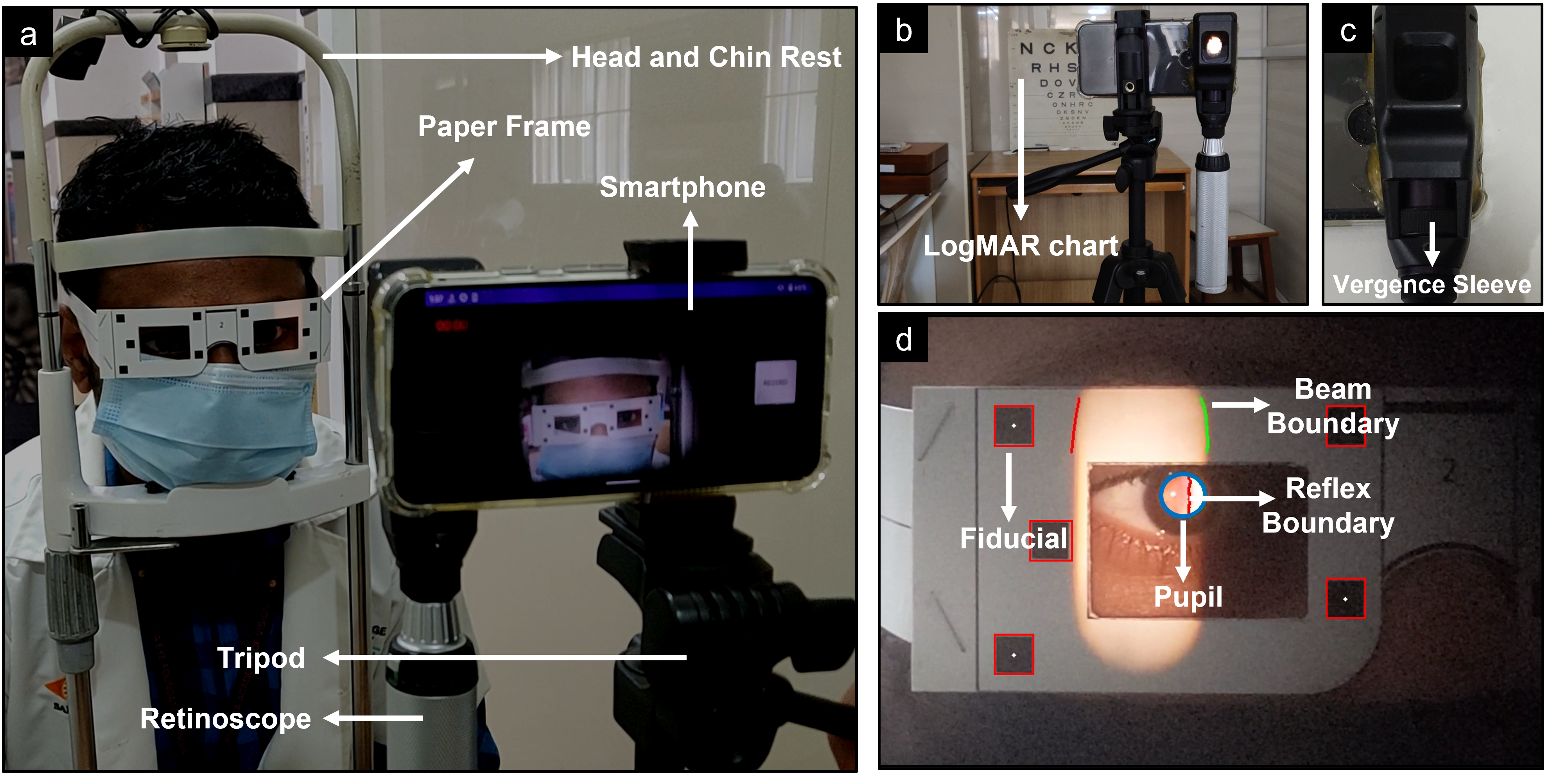}
\end{center}
    \caption{We propose a smartphone-based automatic retinoscopy solution. (a) Our proposed setup consists of a retinoscope attached to a smartphone, (b) the setup from the patient's viewpoint facing the logMAR chart, (c) retinoscope and its vergence sleeve, and (d) a single frame of the digital retinoscopy video with automatic detection of fiducials, pupil, reflex edges, and beam edges, using our video processing pipeline.}
    \label{fig:overview}
\end{figure*}

There are multiple ways to measure refractive error. Subjective refraction is considered the gold standard, which comprises
complex procedural steps performed by an eye specialist. It is subject to the patient's ability to discern changes in clarity while reading the Snellen/logMAR~\cite{logmarChart} chart with a combination of corrective lenses to obtain the best visual acuity. The feedback and cooperation required from the patient make subjective refraction challenging for young children and infants, developmentally delayed adults, or individuals whose behavior limits the ability to be cooperative. Moreover, language barrier between the patient and examiner, and the time required for the iterative process, act as deterrents toward frequent mass screening using subjective refraction in remote areas. 

To address these issues, objective refraction methods (such as retinoscopy, eccentric photorefraction, and autorefractors) have been increasingly used, where no input is required from the patient. Though autorefractors are popular, they are expensive and bulky, and have been found to significantly overestimate myopia~\cite{Choong2006-vy}.
Moreover, they require some cooperation from the patient, making their use challenging for children under 6 years of age~\cite{coopeartive_challenges_AR}. 
In contrast, retinoscopy is highly accurate and requires minimal patient cooperation. Thus, retinoscopy is preferred over autorefractors~\cite{retino_vs_ar_1}, and is the recommended starting point for subjective refraction~\cite{aao_comprehensive_pediatric_eye}.
Retinoscopy is performed using a handheld device called retinoscope, which is low-cost and lightweight.
For children, periodic vision screening using retinoscopy helps in detecting amblyopia risk factors, as amblyopia generally develops from birth up to age 7 years~\cite{amblyopia_age}. Moreover, retinoscopy is used to detect scissors reflex for diagnosing keratoconus~\cite{retino_kerato_1}, a serious corneal disease.

Although versatile and accurate, retinoscopy has certain limitations. The first and most important is that it can only be performed by skilled eye specialists. There is an acute shortage of eye specialists in low-income countries~\cite{30_years_trends}, making retinoscopy inaccessible.
Second, retinoscopy depends on the examiner's perception and is hence prone to inter-observer variability~\cite{retinoscopy_precision, PMID:28030881}. 
Third, retinoscopy requires a lens kit of corrective lenses, increasing the cost and bulkiness. 
These factors limit the use of manual retinoscopy for mass screening efforts.

In this work, we propose an automated way of performing retinoscopy using a smartphone attached to the retinoscope, with the patient wearing a custom pair of paper frames. 
Our proposed digital approach has a few additional benefits compared to manual retinoscopy. It alleviates the need for a lens kit or skilled examiner, and removes subjectivity from the process, thus helping to democratize retinoscopy for large-scale screening. Moreover, digital retinoscopy allows pausing, (re)playing, and digitally zooming into the recorded retinoscopy video, enabling it to be used as a training device to teach retinoscopy. Finally, the digital record can be used to track the progress of abnormal retinal reflex movements.

To estimate eyes' refractive error, we extended retinoscopy's mathematical model (derived from the optical principles of streak retinoscopy~\cite{retino_derivation}) to work with videos captured from a smartphone camera.
Our proposed video processing pipeline automatically extracts the variables needed by the mathematical model and uses it to predict the net refractive error of the patient's eye. 
We evaluated the proposed video processing pipeline in real-world settings on 185 eyes, and achieved a 
mean absolute error of 0.75 $\pm$ 0.67D in net refractive power along the horizontal meridian (\textit{net refractive error = spherical + (cylindrical $\times$ sin(axis)$^2$})~\cite{meridional_refractometry}) compared to subjective refraction measurements. We also output a 4-class classification---normal (-1D to 1D), moderate myopia (-4D to -1D), high myopia (<=-4D), and moderate hyperopia (1D to 4D)---based on the estimated net refractive error, and achieved a sensitivity of 91.0\% and specificity of 74.0\% on our dataset. We make the video analysis pipeline and paper frame details publicly available at our \href{https://www.microsoft.com/en-us/research/project/auto-retinoscopy-automating-retinoscopy-for-refractive-error-diagnosis/}{project page}\footnote{\url{https://www.microsoft.com/en-us/research/project/auto-retinoscopy-automating-retinoscopy-for-refractive-error-diagnosis/}}.

The contributions of our work can be summarized as follows:
\begin{itemize}
\item A method to automate retinoscopy using a smartphone attached to a retinoscope, with the patient wearing a paper frame, thus alleviating dependency of retinoscopy on eye specialist and expensive lens kit.
\item An extension of the retinoscopy mathematical model to estimate refractive error from retinoscopic beam position and retinal reflex position from smartphone-captured retinoscopy videos; a video processing pipeline to automate the retinoscopy mathematical model, which takes retinoscopy video as input and provides net refractive error as output.
\item An evaluation of the proposed approach on 185 distinct eyes from 128 patients in a real-world setting, compared with refractive error measurements using subjective refraction, autorefractor, and manual retinoscopy.
\end{itemize}

To the best of our knowledge, ours is the first work that automates the manual process of retinoscopy to enable low-cost, portable, and accurate refractive error estimation without a skilled examiner. 
We envision that our proposed method and findings will help the community for future research in this direction.

%% file: sections/2_background.tex
\section{Background} \label{sec:background}
Refractive error comprises three components: spherical error, cylindrical error, and cylindrical axis. Under normal circumstances human eye is spherical in shape, thus refractive errors can be corrected solely using spherical lenses, which refract light in all directions. However, in an aspheric eye, called astigmatism, light is refracted differently at different meridians due to the misshapen cornea.
In such cases, cylindrical lenses are prescribed, which 
are curved along the cylindrical axis and flat along the perpendicular axis. 
While the focus of a spherical lens is a single point, the focus of a cylinder lens is a line. 
The cylinder axis is expressed in degrees between 0 and 180.
Most prescriptions have some combination of spherical and cylinder lenses. 
For the purpose of screening, we focus on computing net refractive error, which is a combination of spherical and cylindrical error, along a particular meridian.
Here, we describe the principle of retinoscopy for computing spherical error.

In retinoscopy, the examiner shines a linear streak of light (referred to as the retinoscopic \textit{beam}, Figure~\ref{fig:overview}) and \textit{scopes} (move along a particular axis) the beam on the patient's eye using a retinoscope device from a fixed distance (referred to as the \textit{working distance}). The light beam passes through the different refractive surfaces of the eye, including cornea, anterior chamber, eye lens, and posterior chamber, to form an image (referred to as the \textit{retinal reflex}) on the retina. After reflection from the retina, the light travels back through the same refractive surfaces in reverse order to form another image, which is viewed by the examiner through a peephole and a semi-silvered mirror in the head of the retinoscope. The examiner observes four characteristics of the reflex movement--size, brightness, direction, and speed--and tries to achieve \textit{neutralization} by adding corrective lenses.
The speed of reflex aids in identifying if the eye is close/far from neutralization~\cite{corboy2003retinoscopy}. If the examiner observes a slow-moving reflex, it means the eye is far from neutralization, and as corrective lenses are added, the reflex moves more rapidly.
Similarly, with respect to size, a narrow streak of reflex means the eye is far from neutralization, and as corrective lenses are added, the reflex broadens, finally filling the pupil at the neutralization. With respect to brightness, large refractive errors form a dull reflex, while the reflex becomes brighter as the refractive error reduces. 
The direction of reflex movement is categorized as follows:

\begin{itemize}
\item \textbf{Against Movement}: In this case, the reflex moves in the opposite direction to the retinoscopic beam. The reflected rays from the retina meet at the far point~\footnote{The far point of the eye is the maximum distance at which the eye can see the objects clearly. The far point of the normal human eye (emmetropic eye) lies at optical infinity (6m or more) in the relaxed state.} located between the retinoscope and the patient's eye, and the examiner uses negative lenses to correct this case of myopia.

\item \textbf{With Movement}: Here, the reflex moves in the same direction as the retinoscopic beam. The far point is either located behind the retinoscope or a virtual far point is formed behind the patient's eye. It is corrected by using positive lenses.

\item \textbf{Neutralization}: This is the end goal of retinoscopy, wherein the reflex completely fills the patient's pupil and is independent of the retinoscopic beam movement. At neutralization, the far point of the patient's eye is located at the peephole of the retinoscope. The corrective lens power and the working distance can then be used to determine the eye's refractive error.
\end{itemize}

In retinoscopy, instead of determining the actual far point of the eye,
neutralization shifts the far point to the known position of working distance. The power of the corrective lenses to achieve neutrality is termed \textit{gross power} ($P_{gross}$). The \textit{net refractive power} ($P_{net}$) of the eye can then be computed by adding a correction term for the known working distance (d) as: $P_{net} = P_{gross} - 1/d$.

The examiner might perform \textit{cycloplegic refraction}, in which eye drops are added to the patient's eye to relax the ciliary eye muscles in charge of focusing to reduce accommodation\footnote{Accommodation refers to the ability of eye to change its focus from distant to near objects by chaning the lens shape via ciliary muscles}. This causes the eye lens to be relaxed, making the eye more hyperopic. Hence a cycloplegic correction is applied, depending upon the used eye drops.

Most retinoscope devices provide two controls---a \textit{streak rotator} to control the orientation of the beam (from 0-179$^{\circ}$), and a \textit{vergence sleeve} to adjust the streak from convergent (concave mirror) to divergent (plane mirror).
To perform objective refraction, examiners mainly use diverging light (plane mirror).
Orientation of the beam is used to identify the axis of the cylindrical power (in the case of astigmatism).

In this work, we capture retinoscopy videos by moving the vertical streak of the beam along the horizontal meridian of the patient's eye.
The captured video is passed through a video processing pipeline, which estimates the working distance between the patient's eye and the smartphone, tracks the retinoscopic beam and the position of reflex to output the net refractive error of the eye along the horizontal meridian.

%% file: sections/3_relatedWork.tex
\section{Related Work} \label{sec:relatedWork}

Several portable devices have been proposed over the last few decades for estimating refractive error using a variety of principles, such as self-subjective refraction, eccentric photorefraction, and wavefront aberrometry (Table~\ref{tab:summary_compiled}). Our method, on the other hand, is based on the principle of retinoscopy.
In this section, we focus on prior work related to refractive error estimation using other principles and demonstrate the relevance of our proposed automatic retinoscopy-based solution.

\subsection{Self-subjective Refraction}
Subjective refraction, the gold standard for vision screening, relies on an eye specialist. To reduce this dependency, a few works tried estimating refractive error using the patient's judgment of the sharpness/blurriness of a test object, perceived via novel interactive methods. One of the earliest works, NETRA~\cite{Netra}, utilized an array of microlens placed over a smartphone display and asked the user to interactively align patterns as seen from a pinhole plane on the display. Later, using the same principle, they introduced NETRA autorefractor, a \$1290 commercially-available device that gamifies the process of measuring refractive error via a series of interactions in a virtual-reality environment. A clinical study with 152 eyes found the NETRA device to estimate refractive error accurately (mean absolute error 0.69 $\pm$ 0.53D)~\cite{netra_clinical}. However, a recent work~\cite{netra_time} reported that using the NETRA autorefractor had a steep learning curve for healthcare practitioners and patients alike, and took roughly 10 minutes per patient.

Similarly, Folding Foropter (FoFo)~\cite{fofo} is a $\sim$\$1 paper-based self-refraction tool that aids users in determining their refractive error. It consists of a pair of retractable paper tubes with a lens at each end. The user looks through the paper tubes at a distant target (3m) and adjusts the tube such that the target becomes sharp and in focus. A scale embedded on the sides of the tube depicts the spherical refractive error. USee~\cite{usee} is another self-refraction tool. It comprises of a glass frame with an adjustable progressive lens system for each eye placed in front of a 3mm rectangular slit. An additional refraction bar is moved up and down to change optical power via a dial on each side of the frame until the characters in the logMAR chart are visible clearly. It can estimate spherical refractive power from -6D to +6D. Though FoFo and USee are affordable solutions, they have not been evaluated in a real-world setting. Moreover, the accuracy of self-subjective refraction devices relies on the patients' cooperation and feedback, making it challenging for young children, infants, and developmentally delayed adults.

\input{tables/summary_compiled}

\subsection{Eccentric Photorefraction}
Several objective refraction devices based on the principle of eccentric photorefraction~\cite{eccentric_photorefraction_theory} have also been proposed. The solutions vary from expensive commercial devices with sophisticated hardware (like Plusoptix A09, SPOT vision screener) to low-cost smartphone-based approaches~\cite{smartphone_1_photorefraction, smartphone_2_photorefraction}. These devices estimate refractive error as a function of the size and tilt of the crescent formed on the patient's pupil when illuminated with (infrared/visible) light from a known working distance. The Plusoptix\footnote{Plusoptix: \url{https://www.plusoptix.com/en-us/products}} and SPOT vision screener\footnote{Welch Allyn
Spot Vision Screener: \url{https://www.hillrom.com/en/products/spot-vision-screener/}} have been clinically evaluated and are used to detect myopia, hyperopia, astigmatism, and amblyopia in the pediatric age group~\cite{plusoptix_eval, spot_clinical}. Recently, researchers have utilized image processing and machine learning techniques in conjunction with eccentric photorefraction to estimate refractive error from eye images captured by a smartphone. \citet{smartphone_1_photorefraction} proposed a data-driven approach to estimate refractive error using SVR trained on handcrafted features (like iris radius, pupil radius, and retinal reflex width). In a follow-up work, \citet{smartphone_2_photorefraction} proposed a CNN-based solution to predict refractive error using eye images captured by a smartphone.

The unique advantage of these solutions is that they
require minimal cooperation from the patient, making them ideal for vision screening in schools and pediatric hospitals. However, the eccentric photorefraction principle has a few inherent limitations, limiting its widespread adoption
and questioning the reliability of devices based on this principle.
First, eccentric photorefraction based devices have been found to be inaccurate in predicting refractive power of specific ethnic groups (especially Caucasian, East African, African, and Indian)~\cite{eccentric_ethnicity_challenge} due to variations in defocus calibration factor. Second, such devices accuracy is dependent on illumination distribution around the tear film, which in turn is dependent on the time when the eye image was clicked after a blink.
~\citet{eccentric_simulation} reported significant differences in refractive error estimation across multiple eccentric photorefraction images captured at different time lag after a blink.
Third, the eccentric photorefraction principle's accuracy and operating range rely on the distance between the camera and the light source~\cite{eccentric_photorefraction_slope}, suggesting that the accuracy of smartphone-based approaches~\cite{smartphone_1_photorefraction, smartphone_2_photorefraction} is a function of smartphone-specific parameters.

\subsection{Wavefront Aberrometry}
Devices based on the wavefront aberrometry principle project a wavefront of light on the patient's pupil and record the reflected light from the retina as it passes through the different refractive components of the eye, including the cornea and the eye lens. By analyzing the distortions in that reflected wavefront, a topography map of the eye is created using Zernike decomposition. QuickSee~\cite{Quicksee_AR_evaluation} by PlenOptika is a binocular handheld autorefractor that uses this principle.
\citet{sv_one_AR} leveraged the same principle to create SVOne autorefractor. It uses an expensive hardware module to direct light onto the patient's eye, and a smartphone's camera to capture the reflected light. Apart from estimating refractive error accurately, devices based on wavefront aberrometry can also measure higher-order aberrations in the reflected wavefronts, potentially useful for estimating spherical aberrations and secondary astigmatism. However, these devices are expensive, restricting their access in low- and middle-income countries.

~\\
To summarize, 
commercial vision screeners (like Plusoptix, SPOT, QuickSee, Eye Netra, and SVOne AR) though clinically evaluated, are expensive and require a minimum skill level to obtain reliable measurements. On the other hand, low-cost solutions~\cite{smartphone_1_photorefraction, smartphone_2_photorefraction, usee, fofo} have limited evaluation with respect to refractive error range and/or age group. Most importantly, none of the proposed solutions rely on the retinoscopy principle, which requires minimal patient cooperation, and is considered the best starting point for subjective refraction~\cite{retino_vs_ar_1}. Perhaps, the only digital retinoscopy prior work is by \citet{digital_retino}. They created a digital retinoscope by attaching a standard retinoscope to a smartphone camera, primarily to teach beginners retinoscopy.
However, they neither propose a method to estimate refractive error from the recorded videos nor conduct any evaluation.
Therefore, we believe that our current work is the first step towards automating retinoscopy.

%% file: tables/summary_compiled.tex
\begin{table}[!tb]
\centering
\caption{Summary of approaches for estimating refractive error based on different principles, in contrast to our retinoscopy-based approach. Note: For smartphone-based approaches, `Extra Hardware' states hardware required for diagnosis apart from the smartphone; `Cost' of the solution is categorized into: Low (<= \$500), Medium (\$500 -- \$5000), and High (>= \$5000).}
\resizebox{\textwidth}{!}{
\begin{tabular}{| l | c | c | c | c | c | c | c |}
\toprule
\textbf{Principle} & \textbf{Prior work / Device} & \textbf{Phone} & \textbf{Extra hardware} & \textbf{Cost} & \textbf{Input} & \textbf{Output} & \textbf{Dataset} \\ 
\hline
\hline
{\multirow{3}{*}{\begin{tabular}[c]{@{}l@{}}Self-subjective\\Refraction\end{tabular}}} & \citet{netra_clinical} & Yes & Eye-Netra AR & Medium & User feedback & Sph, Cyl, Axis & 152 eyes \\ \cline{2-8}
& USee~\cite{usee} & No & Lens kit & Low & User feedback & Sph & 120 eyes \\ \cline{2-8}
& FoFo~\cite{fofo} & No & - & Low & User feedback & Sph & - \\
\hline
\hline
{\multirow{4}{*}{\begin{tabular}[c]{@{}l@{}}Eccentric\\Photorefraction\end{tabular}}} & SPOT~\cite{spot_clinical} & No & - & High & Red reflex image & Sph, Cyl, Axis & 134 eyes \\ \cline{2-8}
& Plusoptix A09~\cite{plusoptix_clinical} & No & Computer & Medium & Red reflex video & Sph, Cyl, Axis & 64 eyes \\ \cline{2-8}
& \citet{smartphone_1_photorefraction} & Yes & Trial frame & Low & Red reflex image & Sph & 165 eyes \\ \cline{2-8}
& \citet{smartphone_2_photorefraction} & Yes & - & Low & Red reflex image & Sph & 172 eyes \\
\hline
\hline
{\multirow{2}{*}{\begin{tabular}[c]{@{}l@{}}Wavefront\\Aberrometry\end{tabular}}} & QuickSee~\cite{Quicksee_AR_evaluation} & No & - & High & Wavefront images & Sph, Cyl, Axis & 82 eyes\\ \cline{2-8}
& Ciuffreda et al.~\cite{sv_one_AR_clinic} & Yes & SVOne AR~\cite{sv_one_AR} & High & Wavefront images & Sph, Cyl, Axis & 50 people \\
\hline
\hline
Retinoscopy & Ours & Yes & \begin{tabular}[c]{@{}l@{}}Retinoscope\\ Paper frame\end{tabular} & Low & Retinoscopic video & Net refr power & 185 eyes \\
\bottomrule
\end{tabular}
}
\label{tab:summary_compiled}
\end{table}

%% file: sections/4_dataCollection.tex
\section{Data Collection} \label{sec:dataCollection}
To evaluate our proposed mathematical model and video processing pipeline, retinoscopy videos were captured using a Pixel 4A android smartphone with a Heine Beta 200 streak\footnote{\url{https://www.heine.com/fileadmin/accessfiles/en_GB/download/HEINEBETA200LEDRetinoscopeInstructionforUse.pdf}} retinoscope attached to the smartphone camera at the peephole, with the patient wearing a paper frame (Figure~\ref{fig:overview}a). This allows us to record the beam and reflex movement as observed by the examiner during retinoscopy.
Patient volunteers with different refractive powers were recruited from a local eye hospital. 
Below we describe our data collection procedure, including the paper frame design and data collection app. Next, we elaborate on the diversity of the participant pool and the dataset used for evaluation. All details of our study were approved by the hospital's Institutional Review Board (IRB).

\subsection{Data Collection Procedure}
The data was collected from Sept-Dec 2021 in collaboration with a local eye hospital. The eye hospital is a leading eye care and teaching institution in the country, with more than 15 eye doctors and 10 optometrists. 
For each patient who agreed to participate in the study, 
we collected subjective refraction data for both the eyes. Additionally, we collected two sets of medical-grade refractive error measurements, using autorefractor (Topcon KR 8900\footnote{\url{https://topconhealthcare.com/product-category/auto-refractometer}}) and manual retinoscopy, whenever possible.
The data was collected by a group of staff optometrists (25-40 years age, 1-10 years of experience).
Retinoscopy video data from our setup was collected by two student interns (male, 21, and 23 years of age), who were trained by one of the researchers to use the data collection app.

To collect video data, patients were asked to wear a comfortable size of our paper frame. They were seated on a chair with their chin placed on an adjustable chin-rest and were instructed to focus on the LogMAR chart placed at a 3m distance along the eye's optical axis for relaxing the eye to minimize accommodation.
After entering the patient's demographic information (gender, age) and patient ID on the data collection app, the app prompts the data collector to record the video for the left eye, followed by the right eye.
For each eye, the data collector needs to rotate the retinoscope attached to the smartphone such that the beam moves from one end of the paper frame to the center and back.
This is referred to as a \textit{pass}. This process is repeated to obtain four passes (2 left-to-right and 2 right-to-left) for a single eye in a video.
The retinoscope-attached smartphone was mounted on a tripod stand to minimize camera jitter while panning the retinoscopic beam across the eye. 
Several instructions were given to the data collector---vergence sleeve should be at the lowest point to ensure plane mirror effect; to adjust chin-rest height such that the patient's eye is at the same level as the LogMAR chart; to pan the device slowly from left-to-right and back in order to capture more video frames; ensure that the beam is vertical (90$^{\circ}$) using the streak rotator, etc.

The hospital staff helped us in recruiting patients as participants.
Patients with cataracts or active signs of ocular infection/conjunctivitis/acute eye trauma were excluded from our evaluation. Since retinoscopy-based measurements are possible only when the reflex is visible in the eye, patients with dull reflexes due to corneal haze or very high myopia and small pupil size (less than 3mm) were also excluded. 
During data collection the participants were given an IRB-approved consent form to read and sign before their data was collected.
Every evening, the data collector connected the phone to the hospital's WiFi network to auto-sync the collected videos along with the metadata to a cloud-based storage.
Refractive error readings from autorefractor, retinoscopy and subjective refraction, were recorded from the hospital's database by matching the patient's ID saved during data collection. Note: We use the gold standard subjective refraction as the ground truth. Participants were not paid to participate in the study. It took 5-7 minutes per participant for video data collection from our app.

\subsection{Paper Frames}

Computation of refractive error using the retinoscopy mathematical model (Section~\ref{sec:derivation}) needs the position of beam and reflex w.r.t a \textit{fixed} world co-ordinate system. However, since in our setup the video is captured via a moving camera, we need a fixed target that can be used as a reference point. For this, we use a custom pair of paper frames (without any lens) with a unique pattern of known dimensions (Figure~\ref{fig:glasses})
as a robust and low-cost solution. The glasses also allow us to compute the working distance between the smartphone camera and eye accurately, which is again needed by the retinoscopy mathematical model.
A single size of glasses for patients with different facial sizes posed a challenge, thus we created three paper frame sizes with varying reflex search space dimensions (width $\times$ height)---(1) small: 3.0cm $\times$ 2.0cm, (2) medium: 3.5cm $\times$ 2.0cm, (3) large: 4.0cm $\times$ 2.0cm.
We further number-coded them for easy identification. We tried a variety of unique patterns to estimate the working distance accurately and track the retinoscopic beam (which is falling partially on the top part of the paper frame). Here we describe three patterns and the issues we identified.

\subsubsection{Colored Stripes}
Inspired by~\cite{BiliScreen}, we tried colored stripes of fixed width along the edges of the frame (Figure~\ref{fig:glasses}a).
For processing the video frames, we leveraged color thresholding and known strip width to uniquely identify and segment the colored stripes. However, the fixed color threshold values failed to detect the colored stripes robustly in varying ambient lighting of the room.

\subsubsection{BW Stripes with Square Fiducials}
Instead of multiple colored stripes, we switched to black-and-white stripes (Figure~\ref{fig:glasses}b) as it needed only a single color threshold value. We also added black squares of fixed known size at the corners (similar to QR code corners) as fiducial markers. The size of fiducials helped us to estimate the working distance, and the stripes helped us to track the retinoscopic beam. However, the processing failed if even a single fiducial marker among the three was undetected due to glare, cropping, or any error. Moreover, the beam light intensity interfered with the stripes, resulting in erroneous beam edge detection.

\subsubsection{Square Fiducials}
\begin{figure*}
\begin{center}
    \centering
    \includegraphics[width=1\linewidth]{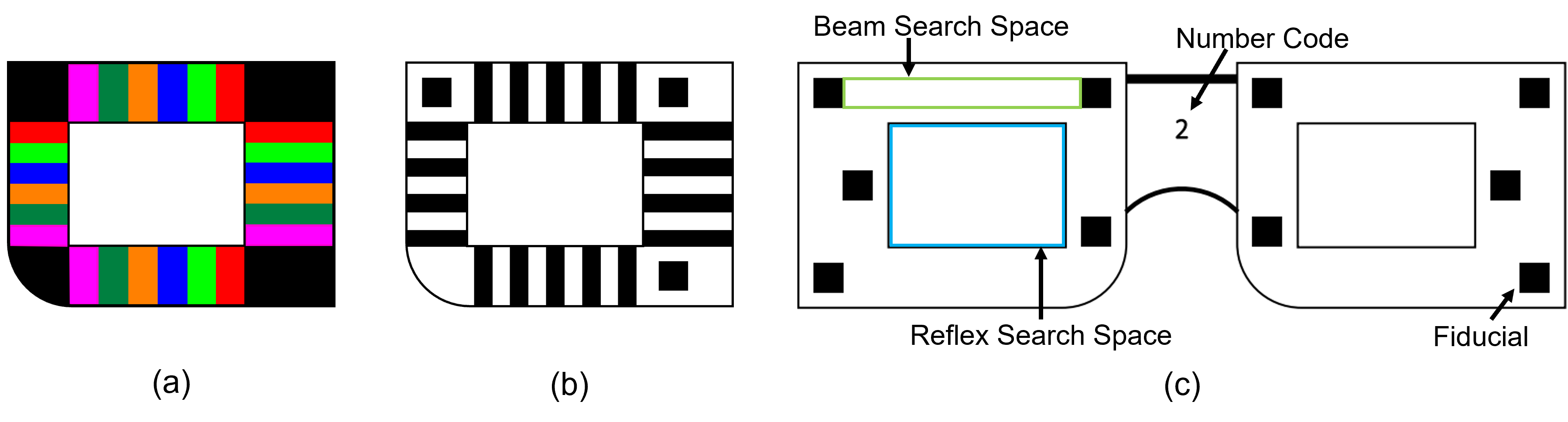}
\end{center}
    \caption{Different paper frame designs we tested: (a) frame with colored stripes, (b) frame with black-and-white stripes and square fiducials, and (c) the final frame design with five fiducials.}
    \label{fig:glasses}
\end{figure*}

We removed the black-and-white stripes and defined the empty space between fiducials as the search space for beam edges (Figure~\ref{fig:glasses}c). We added two more fiducials along the vertical sides of the frame to ensure that the video processing is robust against
missing fiducial(s). The fiducials were placed such that the relative position of any fiducial w.r.t any other fiducial is unique. During the video processing, this allowed us to uniquely identify the detected fiducials if any two (out of 5 for each eye) fiducials were detected. This made the frame localization step of our video processing pipeline very robust.

\subsection{Data Collection Parameters} \label{sec:dataCollectionParameters}
During the initial stage of data collection, we observed that external factors (like room brightness, working distance), video recording settings (like FPS, video resolution), and retinoscope settings (like vergence sleeve position) significantly affected the captured video quality.
Hence, we followed an iterative approach to identify their impact on our video processing pipeline and recommend the most suitable setting.

\subsubsection{Room brightness}
Generally, retinoscopy is performed in a dark setting, as it increases the pupil size to allow more light to enter the eye, enabling the examiner to observe a clear reflex movement. 
Hence we followed a similar dark room setting for capturing retinoscopy videos. However, videos in the low-light setting are grainy, and at times, all the five fiducials were not visible. We solve these problems via our paper frame design and by leveraging image processing techniques like median filtering for noise reduction and adaptive histogram equalization for contrast adjustment. Moreover, capturing videos with no light source in the room helped us avoid unwanted reflections in the eye.

\subsubsection{Working distance}
Retinoscopy is performed with the examiner sitting at an arm's length (50 cm to 1 m) away from the patient. This allows the examiner to quickly add/remove corrective lenses on the trial frame. 
As our proposed approach does not require external lenses, 
we used a smaller working distance (approximately 30-40 cm) to obtain a larger resolution in our region of interest (i.e., pupil). Its impact on our proposed mathematical model and results has been discussed in Section~\ref{sec:mathematical model results}.

\subsubsection{FPS and Video resolution}
We had a few options for frames/second (30, 60, or 120 fps). Although higher fps provided more precise tracking of beam and reflex movements, it resulted in noisier video in low-light settings as exposure time decreased with increasing fps. 
Hence we selected 30 fps for recording the videos. 
For video resolution, we selected the highest available option (3840 $\times$ 2160), as an increase in video resolution resulted in capturing more pixels in our region of interest, which helped us in more accurately detecting the beam and reflex position.

\subsubsection{Vergence sleeve position}
In retinoscope device, the position of the light source is fixed, but the position of the condensing lens inside the retinoscope is controlled by the examiner using the vergence sleeve~\cite{corboy2003retinoscopy} (Figure~\ref{fig:overview}c). 
By changing the sleeve position, the retinoscopic beam vergence is altered, thus manipulating its focal point. 
In the sleeve up position, the streak emanates as a converging beam (concave mirror effect), and the beam is focused and very bright, thus making it difficult for the video processing pipeline to track the retinal reflex movement. When the retinoscopy sleeve is lowered, the streak emanates as a diverging beam (plane mirror effect), which the examiners use most often. We also recorded our videos with the plane mirror setting.
\begin{figure*}
\begin{center}
    \centering
    \includegraphics[width=1\linewidth]{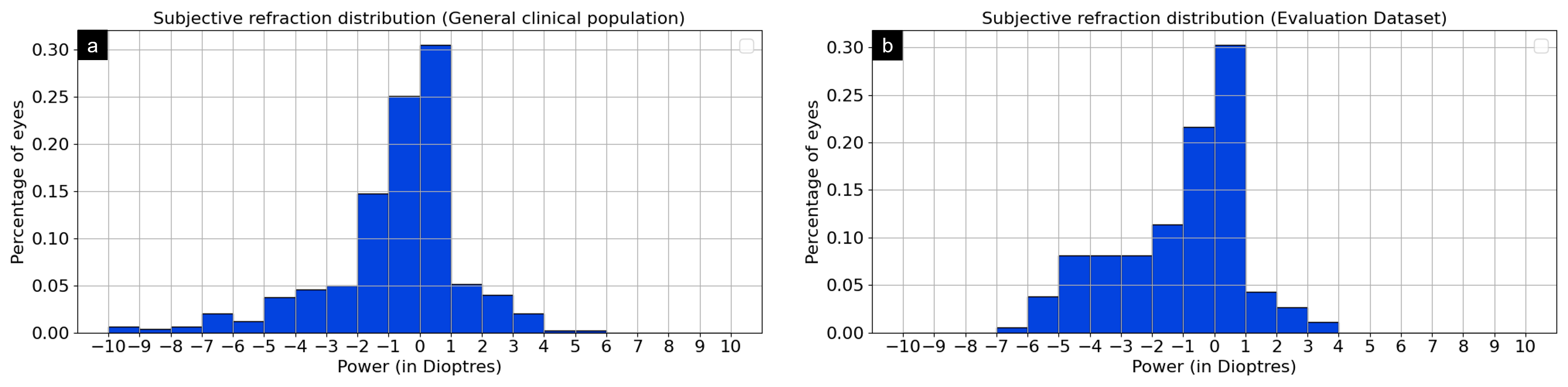}
\end{center}
    \caption{Normalized distribution of the percentage of refractive power (in Dioptres) for (a) historical data collected at the local eye hospital, and (b) our dataset. Both show a similar distribution with majority of samples ($\sim$95\%) lying between -6D to 4D.}
    \label{fig:distribution}
\end{figure*}

\subsection{Data Collection App}
To collect data, we developed an Android application wherein,
for each patient, the data collector enters the age, sex, and hospital-assigned patient ID. The patient ID is the only link between the captured video and the medical-grade refractive error measurements (using autorefractor, retinoscopy, and subjective refraction methods) of the patient. Next, the data collector has an option to choose data collection parameters, such as working distance, vergence sleeve position, fps, and video resolution. Our recommended values for these parameters are selected by default.
The app also allows the data collector to select the eye drops used in case of cycloplegic refraction. This helps our refractive error estimation algorithm to compensate for the accommodation due to the cycloplegic effect. 
Next, the app prompts the data collector to start recording video for the left eye. After recording, the app provides an option to recapture the left eye, start recording video for the right eye, or end the session. 
After completing the session, the app asks the data collector to enter refractive error measured using an autorefractor, retinoscopy, and/or subjective refraction method for both eyes. For each patient, the app stores a metadata file containing the patient's demographic information, data collection parameters, medical-grade refractive error, etc. The collected data is auto-synced to cloud-based storage in a best-effort manner. The videos were recorded using a Google Pixel 4A smartphone, utilizing its primary camera (12MP, f/1.7 aperture, 4.4mm focal length, 5.6mm $\times$ 4.2mm sensor size).

\subsection{Participants and Dataset}
128 patients (74 male, 54 female) with an average age of 30.2 $\pm$ 11.1 years (7-58 years age range) volunteered for the data collection. (Note: Due to COVID-19, we were able to recruit very few patients, resulting in a relatively small dataset). Out of the 256 eye videos collected (128*2), 71 videos were removed due to these factors: (1) retinoscope device movement was very fast, (2) out-of-focus/blurry video, (3) patient was blinking throughout the session, and (4) very small pupil size (happened mostly with old-age patients).
Our final dataset comprises 185 distinct eyes (12 dilated and 173 non-dilated eyes), for which we have both smartphone-captured retinoscopic video data and ground truth subjective refraction measurements. The subjective net refractive error along the horizontal meridian ranged from -6.25D to 3.23D (Figure~\ref{fig:distribution}b shows normalized distribution), with 74 myopic (<-1D), 15 hyperopic (>1D), and 96 normal (-1D to 1D) eyes.
Among these 185 eyes, we have autorefractor measurement data for 161 eyes and retinoscopy measurement data for 130 eyes.

As we found our dataset refractive error distribution to be skewed towards myopic eyes, we also collected all patients data (autorefractor, retinoscopy, and subjective refraction measurements for right eye) for 2 days from the hospital's database, to compare the refractive error distribution of our dataset with the general clinical population. In 2-days, 506 patients (255 male, 251 female) with mean age 40.7 $\pm$ 21.7 years had undergone subjective refraction, and the mean spherical power was -0.42 $\pm$ 2.58D. Comparing the normalized distribution of subjective power for the general clinical population (Figure~\ref{fig:distribution}a) with our dataset (Figure~\ref{fig:distribution}b), we found the Kullback-Leibler divergence~\cite{KLD} to be 0.1, thus it is evident that our dataset is a good representation of refractive error among the clinical population.

%% file: sections/5_retinoscopyDerivation.tex
\section{Optics behind retinoscopy}
\label{sec:derivation}
In retinoscopy, the examiner observes the speed, direction, size, and brightness of the retinal reflex to achieve neutralization. In this section, we describe the optical principle behind manual retinoscopy, its mathematical formulation only using speed and direction of beam and reflex~\cite{retino_derivation} (note: the formulation encapsulates information provided by the reflex size and brightness~\cite{retino_derivation}), and extend it to estimate refractive error from smartphone-captured retinoscopy video.

\subsection{Mathematical Modelling}
\begin{figure*}
\begin{center}
    \centering
    \includegraphics[width=1\linewidth]{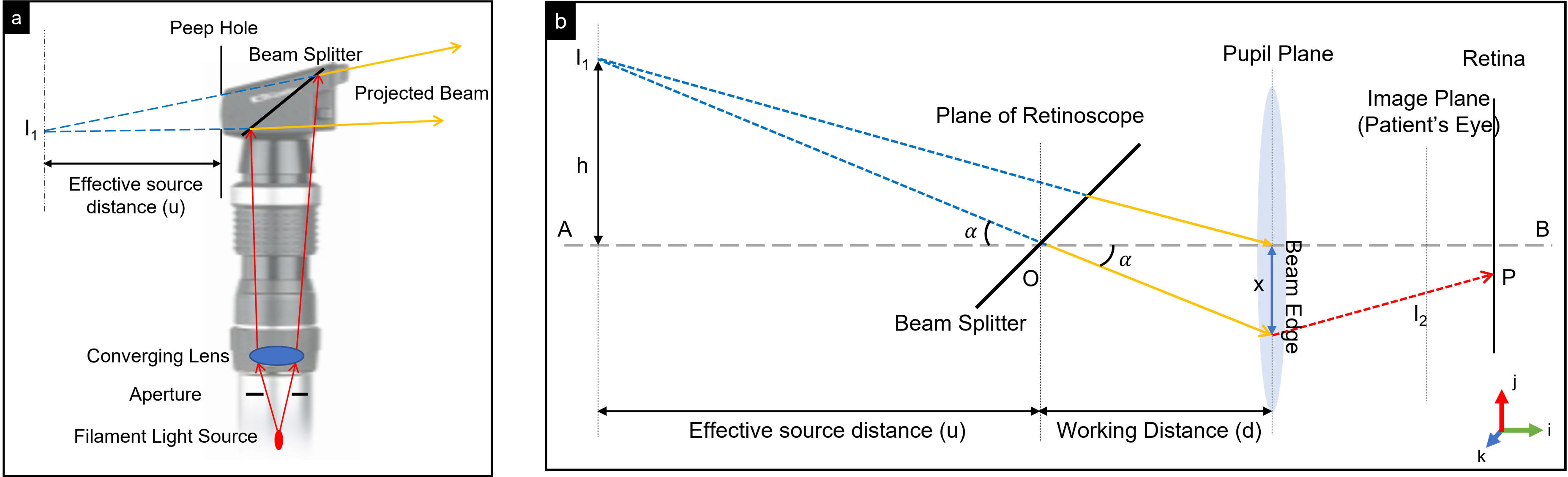}
\end{center}
    \caption{Retinoscopy working principle. (a) Different parts of a retinoscope device, and (b) ray-diagram for a case when the retinoscope is rotated about $\hat{k}$ axis by an angle $\alpha$.}
    \label{fig:ray_diagram}
\end{figure*}

In streak retinoscopy, light from a linear filament bulb passes through a small aperture via a strong converging lens (Figure~\ref{fig:ray_diagram}a). (Note: The bulb and lens are located in the handle of the retinoscope.) The position of the lens with respect to the bulb is controlled by the examiner using the \textit{vergence sleeve}, which changes the vergence of the projected beam. A beam splitter at a 45$^{\circ}$ angle present at the top of the retinoscope, changes the direction of light by 90$^{\circ}$ towards the patient's eye. Extrapolating these rays backwards produces a virtual image ($I_{1}$) which acts as a light source for the patient (Figure~\ref{fig:ray_diagram}a). However, only the rays travelling from $I_{1}$ through the peephole form a beam and reach the patient's eye. We refer to the distance between $I_{1}$ and retinoscope as the effective source distance ($u$; which is a retinoscope specific parameter). 
We captured videos using a Heine Beta 200 retinoscope at the lowest sleeve position (i.e., $u=40cm$).

During retinoscopy, an examiner rotates the retinoscope about the $\hat{k}$ axis from a fixed working distance $d$ (Figure~\ref{fig:ray_diagram}b), where working distance is defined as the distance between the apex of the eye and the camera. The retinoscope's rotation moves the beam along the $\hat{j}$ vertical axis on the pupil plane, sweeping a small angle with the optical axis $AB$, and shifting the virtual light source vertically. For example, Figure~\ref{fig:ray_diagram}b shows the beam at an angle $\alpha$, and the virtual light source ($I_{1}$) shifted by a distance \textit{h} above the optical axis $AB$.
The rays from the virtual light source $I_{1}$ exit the peephole of the retinoscope in the form of a beam, and enter the patient's eye through the pupil at a distance \textit{x} from the optical axis $AB$.
After entering the eye, light undergoes multiple refractions through the cornea, anterior chamber, eye lens and the posterior chamber. For simplicity, we assume all those refractions to take place at the \textit{pupil plane}. These rays then form an image $I_{2}$ (Figure~\ref{fig:ray_diagram}b) on a plane, whose position is dependent on the refractive power of the patient. Note: For emmetropic eye (i.e., eye without any refraction error), the image plane will coincide with the retina; for myopic and hyperopic eye, the image plane will be before and after retina, respectively\footnote{This is for the \textit{gross power}, $P_{gross}$ before adding the working distance correction.}. 
However, since $I_{2}$ is not directly visible to the examiner, it cannot be used to estimate the refractive error. The rays finally illuminate a region on the retina at point P.

\begin{figure*}
\begin{center}
    \centering
    \includegraphics[width=1\linewidth]{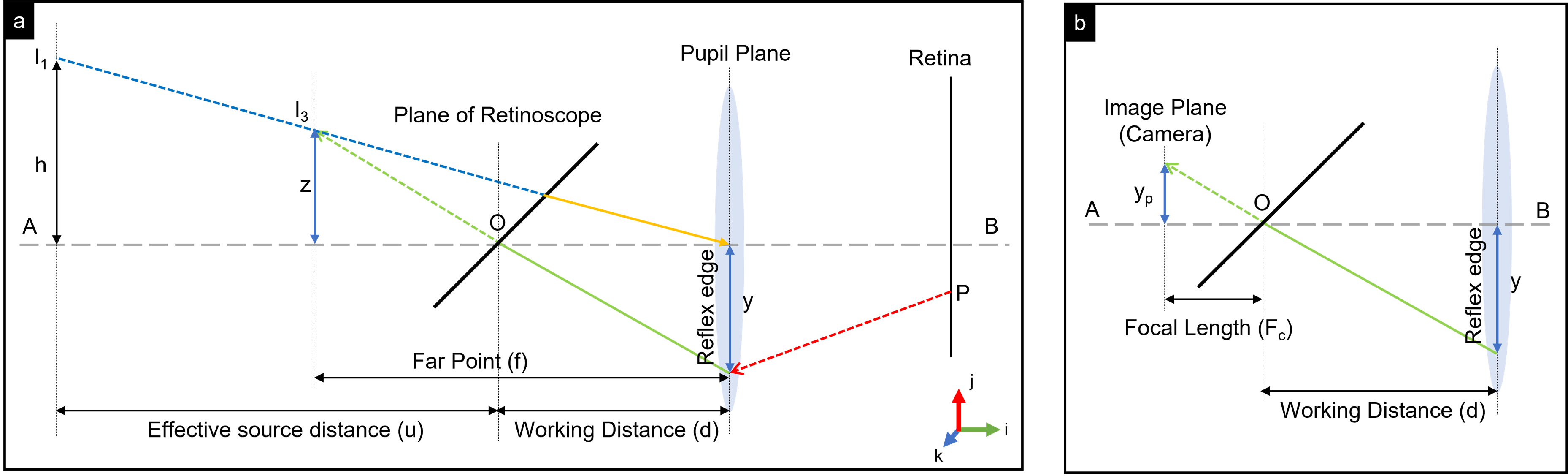}
\end{center}
    \caption{Mathematical modelling of the retinoscopy principle to estimate the refractive power of a given eye $P_{net}$ as a function of effective light sources distance $u$, working distance $d$, and the ratio $y_p$/$x_p$ (= $r$), where $y_p$ is the image location of reflex and $x_p$ is the image location of beam on the camera sensor.}
    \label{fig:ray_diagram2}
\end{figure*}

After the diffused reflection from retina, point P acts as a light source
(Figure~\ref{fig:ray_diagram2}a), and rays emanating from $P$ undergo multiple refractions to emerge out of the pupil. A subset of these rays, at distance $y$ from optical axis $AB$, pass through the peephole of the retinoscope into the examiner's eye, creating a perception of the presence of a bright region in patient's pupil (referred as the reflex). These emergent rays finally meet and form an image $I_{3}$ at the \textit{far point}, at a distance $f$ from the pupil plane.
For illustration purposes we only show the central ray emerging from the light source and subsequent images.

Based on the similar triangles in Figure~\ref{fig:ray_diagram}b:

\begin{align}
    \frac{h}{x} = \frac{u}{d}
    \label{eq:similar_triangles_1}
\end{align}

Similarly in Figure~\ref{fig:ray_diagram2}a:

\begin{align}
    \frac{z}{h} = \frac{f}{u + d} \\
    \frac{y}{z} = \frac{d}{f - d}
    \label{eq:similar_triangles_2}
\end{align}

Multiplying these 3 equations we get:

\begin{align}
    \frac{y}{x} = \frac{u * f}{(f - d) * (u + d)}
    \label{eq:final_res1}
\end{align}

In equation~\ref{eq:final_res1}, the values x and y represent the position of beam and the reflex, respectively, in the 3D world co-ordinate system. A version of this formulation has been proposed earlier in ~\cite{retino_derivation}.
To extend this formulation for refractive error estimation using smartphone captured videos, we further model the formation of the reflex and beam's image on the smartphone camera image plane. Figure~\ref{fig:ray_diagram2}b shows the image formation on a camera of focal length $F_{c}$, where the reflex at a distance $y$ from the optical axis $AB$ forms an image at $y_{p}$ on the camera's image plane. The beam at position $x$, will similarly form a corresponding camera image at $x_p$. Again, from similar triangles:

\begin{align}
    \frac{x}{d} = \frac{x_{p}}{F_{c}};
    \frac{y}{d} = \frac{y_{p}}{F_{c}}
    \label{eq:camera_space}
\end{align}

Substituting $y/x$ from equation~\ref{eq:camera_space} in equation~\ref{eq:final_res1}:

\begin{align}
    \frac{y_{p}}{x_{p}} = \frac{u * f}{(f - d) * (u + d)} \\
    y_{p} * (f - d) * (u + d) = u * f * x_{p}
    \label{eq:camera_space_result}
\end{align}

Note that in this formulation $x_p$ and $y_p$ are measured from the optical axis ($AB$) of the pupil, which is not known.
To get around this, we compute $y_{p}$ and $x_{p}$ at two different timestamps $t_{1}$ and $t_{2}$, and take their difference. In equation~\ref{eq:camera_space_result}, $u$ and $d$ are fixed for a patient during a retinoscopy session, and so is the far point distance $f$, as it is a property of the patient's eye. Thus, computing  equation~\ref{eq:camera_space_result} at times $t_{1}$ and $t_{2}$, and take the difference yields:

\begin{align}
    [(y_{p})_{t_{2}} - (y_{p})_{t_{1}}] * (f - d) * (u + d) = u * f * [(x_{p})_{t_{2}} - (x_{p})_{t_{1}}] \\
    \Delta y_{p} * (f - d) * (u + d) = u * f * \Delta x_{p},
    \label{eq:timestamps}
\end{align}
where $\Delta y_{p} = (y_{p})_{t_{2}} - (y_{p})_{t_{1}}$ and $\Delta x_{p} = (x_{p})_{t_{2}} - (x_{p})_{t_{1}}$ are referred to as the displacement of the reflex and the beam between timestamps $t_{1}$ and $t_{2}$. 

Next, we estimate the refractive error in terms of the far point distance $f$. Let the eye's optical system be represented by a single lens at the pupil plane with the retina at a distance $R$ from the pupil plane. For an emmetropic eye (i.e., eye without any refraction error), the eye should have a refractive power $P_{e}$, such that rays from infinity focus on the retina. Applying lens equation for this condition yields:
\begin{align}
    P_{e} = \frac{1}{\infty} - \frac{1}{R} = - \frac{1}{R}
    \label{eq:emmetropic_power}
\end{align}
Now, if the retinoscopic reflex is formed at a far point distance $f$, it means that the eye has a refractive power, $P_u$, such that the image of point P at distance $R$ is formed at $f$ (Figure~\ref{fig:ray_diagram2}a). Applying the lens equation for this condition now yields:
\begin{align}
    P_{u} = \frac{1}{f} - \frac{1}{R}
    \label{eq:uncorrected_power}
\end{align}
thus the power of the corrective lens, $P_{net}$, needed to fix such an eye should be such that $P_{e} = P_{u} + P_{net}$. Thus $P_{net} = P_{e} - P_{u} = -\frac{1}{f}$. Finally, substituting $f$ from equation~\ref{eq:timestamps} and replacing $\Delta y_{p} / \Delta x_{p}$ as $r$, we obtain $P_{net}$ as:
\begin{align}
    P_{net} = -\frac{1}{f} = \frac{u - (u+d)*r}{r*d*(u+d)}
    \label{eq:final_result}
\end{align}
At this point $P_{net}$ represents the net refractive power of the eye along the vertical meridian $\hat{j}$. To obtain spherical, cylindrical power and cylindrical axis of the lens, we can repeat this process along two more predetermined axis and use the results from ~\cite{meridional_refractometry}.

%% file: sections/6_imageProcessing.tex
\section{Proposed Solution: Video Analysis Pipeline}
\label{sec:videoProcessing}
Our video analysis pipeline takes as input the recorded video
and the frame size (small / medium / large) used to capture the video, and outputs the net refractive power of the eye along the scoped meridian (i.e., the meridian along which the retinoscope beam was scoped). For estimating refractive power, four data points are needed: (1) camera parameters, (2) pixel location of the fiducial centers, (3) pixels moved by the retinoscopic beam between timestamps $t_{1}$ and $t_{2}$, and (4) pixels moved by the retinal reflex between timestamps $t_{1}$ and $t_{2}$. Only the camera parameters are known from the smartphone specifications. Below we describe the different components of our proposed pipeline to compute the remaining variables. Note: As the video frames are very noisy, we use Otsu's thresholding on different regions of the frame---beam search space, reflex search space, etc.---as Otsu is dynamic and adaptive, and hence is generalizable across video quality and lighting conditions.

\begin{figure*}
\begin{center}
    \centering
    \includegraphics[width=1\linewidth]{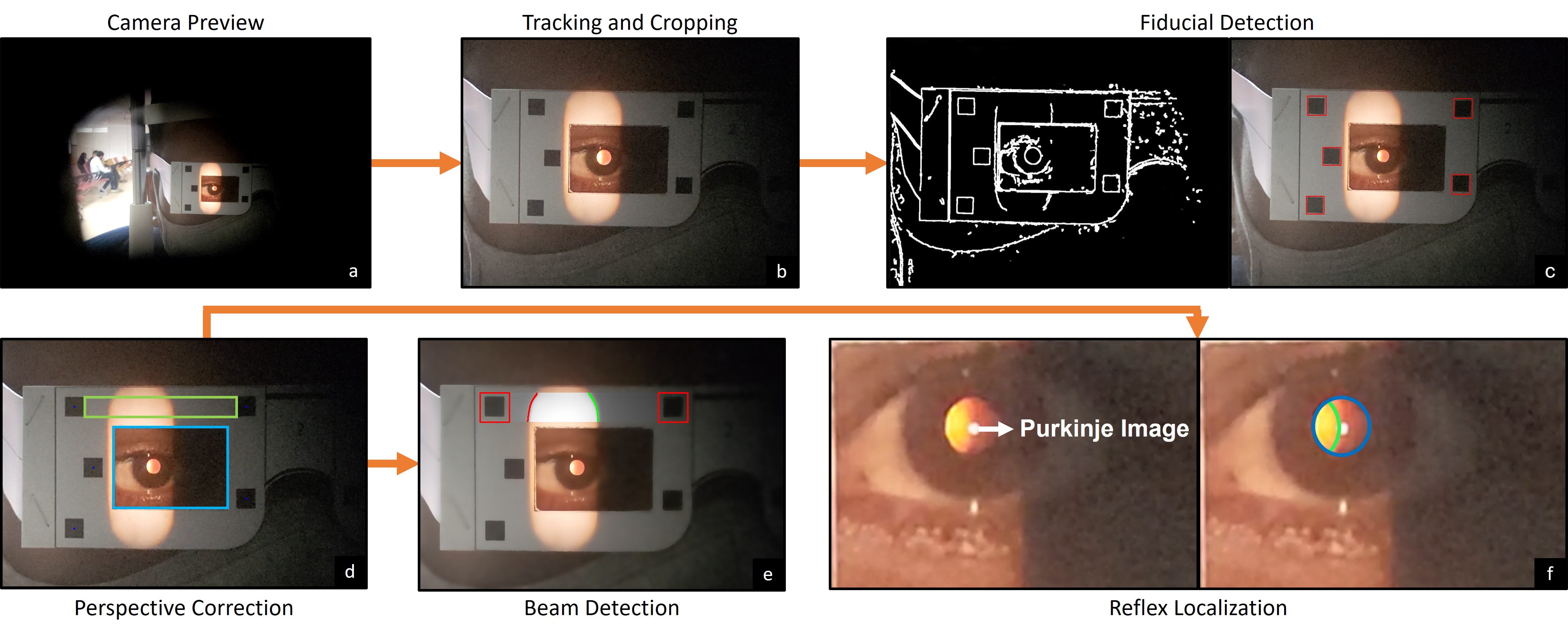}
\end{center}
    \caption{Our proposed video processing pipeline: (a) captured video frame using smartphone camera, (b) frame is initialized with a tracker and cropped to get region of interest, (c) fiducial detection on the edge map, (d) perspective correction via homography estimation, (e) detection of retinoscopic beam edges, and (f) pupil detection and reflex edge detection.}
    \label{fig:pipeline}
\end{figure*}

\subsection{Image Cropping and Tracking}
The first step in our processing pipeline is to define the region of interest (ROI) in each frame of the video. Since the recorded videos are at a 3180 $\times$ 2160 pixels (width $\times$ height) resolution, a significant portion of the frame comprises of the background  (Figure~\ref{fig:pipeline}a), which needs to be removed for subsequent processing. 
Using the fiducials, we detect the paper frame and form a bounding box around it as our ROI  (Figure~\ref{fig:pipeline}b). The ROI is detected on a single frame, and tracked on successive frames using CSRT tracker~\cite{CSRT_tracker}.
In the following steps, we use this cropped ROI ($I_{ROI}$) for each frame as input.

\subsection{Fiducial Detection}
In the cropped region, $I_{ROI}$, the location and size of fiducial squares need to be detected accurately to estimate the working distance. Moreover, location of fiducials also define the search space for beam and reflex detection. 
As the videos were collected in a dark room, there is uneven distribution of light on the paper frame (\textit{e.g.}, the region of the paper frame closer to the retinoscopic beam is more illuminated than other regions) which makes accurate detection of fiducials non-trivial.
We perform three preprocessing steps: (1) convert $I_{ROI}$ to gray-scale based on CCIR-601 standards to obtain $I_{G}$, (2) noise reduction on $I_{G}$ using median filtering with a 5$\times$5 kernel, and (3) adaptive histogram equalization~\cite{CLAHE}.
Next, we obtain an edge map of $I_{G}$ using Canny edge detector~\cite{Canny}. As the square shaped fiducials were captured with a camera panning horizontally, in every frame, a few fiducials appeared as quadrilaterals. So, we identify contours and detect shapes having 4 corners in the edge map. As the fiducial markers would be captured as roughly of same size, we remove quadrilaterals whose area is significantly greater/lesser than median area of the detected quadrilaterals.

To further remove any falsely detected fiducial(s), we leverage the unique fiducial pattern and its dimension to find a fixed ordering among the fiducials. We start with the top-left detected quadrilateral in the frame, and try to fit a set of other detected quadrilaterals using the known fiducial pattern dimension. This step is repeated, starting with each of the identified quadrilaterals, to find the best fit.
If the set of identified fiducials is less than five, the locations of the unidentified fiducial(s) is 
interpolated/extrapolated by tracking their location in consecutive frames.
Due to the low resolution of fiducials ($\approx 40 \times 40 px$), we found aliasing artifacts on the fiducial edges. Hence, we estimate the center of detected fiducials using intensity-based weighted averaging on the negative of $I_{G}$ (\textit{i.e.}, 255 -- $I_G$). 

\subsection{Perspective Correction}
Note that during retinoscopy, participants are asked to look at a logMAR chart placed at a distance of 3m to relax the eye and minimize accommodation. We followed a similar guideline for our data collection. To avoid blocking patient's view of the logMAR chart, we captured our data by keeping the retinsocope at an angle of $\sim$10-15$^{\circ}$ from the patient's line of sight. This resulted in a perspective distortion in the recorded frames.
In this step, we remove this distortion in each frame by homography correction using the detected fiducials.
The fiducial markers on the paper frame lie on a 2D plane in the 3D world. Hence we define a 3$\times$3 homography matrix ($H$) which maps the points from the captured image ($I_{ROI}$) to a fixed known 2D plane~\cite{Hartley2004}. We compute this $H$ matrix by mapping the centers of the five detected fiducials in $I_{ROI}$ to the centers of the known fiducial pattern dimension on the paper frame, using a least square method.
Finally, we apply the transformation $H$ to the image $I_{ROI}$ to 
generate a distortion free image with the correct perspective. 

\subsection{Beam Detection}
We detect the retinoscopic beam edges in the beam search space (Figure~\ref{fig:glasses}) between the top two fiducials. Since the videos are recorded in a low-light setting, 
we used bilateral filtering on $I_{ROI}$ to reduce noise while preserving the beam edges. 
The beam is orange in color, hence 
after cropping the beam search space, we perform non-local mean denoising~\cite{ipol.2011.bcm_nlm} on the red channel of $I_{ROI}$.

There are two possible beam positions. First, the edge is not present in the beam search space. We remove such frames from analysis by comparing the threshold value obtained using Otsu thresholding algorithm~\cite{OTSU} with the empirically found mean intensity of the beam. Second, the beam is partially/fully present in the beam search space. In that case, the search space can be divided into two intensity-based clusters---beam and background. To identify the beam edges, we use Canny edge detector~\cite{Canny} with two threshold values---Otsu threshold and median intensity of the image as threshold---and compute a bitwise OR operation between their outputs. Finally, we label the beam edges as left/right based on the sign of the horizontal gradients at these edges. The gradient value is positive for the left edge as we are moving from low to high pixel intensity and negative for the right edge.

\subsection{Pupil Detection}
The search space for pupil detection is the empty rectangular area in the interior of the paper frame.
We use Hough transform~\cite{HOUGH_1} on the edge map of the cropped reflex search space ($I_{eye}$) to detect the circular pupil. The reflected light from the retina illuminates eye's blood vessels making it appear reddish-orange in color, thus we use the red channel in the RGB color space for pupil detection in the Hough transformed image.

In some frames, Hough transform fails and outputs false positives.
Since the patient focuses on the LogMAR chart during data collection, the position of the pupil with respect to the fiducial pattern remains approximately fixed. Therefore, we create a 2D histogram plot of the x,y coordinates of the Hough-detected circle centers with the top-left fiducial as the origin. The peak in this histogram is selected as the pupil coordinates for that frame.

\subsection{Reflex Edge Localization}
The region of interest for reflex detection is the pupil. Finding the accurate location of the reflex edges (in $I_{eye}$) in each frame is the very challenging, because of the small resolution of the pupil (typically $\sim15$ pixels). As a result, even a single pixel error in reflex edge localization significantly impacts our results.
Therefore, we super-resolve $I_{eye}$ to 4X using FSRCNN (Fast Super-Resolution CNN)~\cite{FSRCNN}, which helps us to obtain a sub-pixel resolution.
Next, we perform Otsu adaptive thresholding~\cite{OTSU} on the super-resolved $I_{eye}$ to localize one/two boundaries of the retinal reflex in each frame.
One of the challenges in the reflex edge localization is the presence of reflections from different surfaces of the patient's eye, referred to as Purkinje images~\cite{Purkinje} (Figure~\ref{fig:pipeline}f). Our video data contains Purkinje image formed by the reflections from the outer surface of cornea, which appears as a bright region, and looks similar to the retinal reflex.
To remove Purkinje reflections in $I_{eye}$, we traverse the binary mask indicating the reflex / not-reflex pixels outputted by the Otsu thresholding, and for each column count the percentage of pixels identified as reflex. Since the Purkinje reflextions are typically much smaller than the reflex, we can discard columns with have only a small percentage of pixels identified as reflex (we used a threshold of 40\% obtained empirically).

Finally, we label the boundaries as left/right edge based on the sign of the gradients along horizontal axis.
Furthermore, we use a gradient-based weighted averaging, where the kernel is a gradient along horizontal axis at that pixel, thus increasing the accuracy of reflex edge localization.

\subsection{Refractive Error Estimation}
After processing all the frames in a video, we obtained five fiducial centers, left and right edge of the beam, pupil center and radius, and left and right edge of the reflex, for each frame. To estimate the net refractive error of the eye according to our formulation~\ref{eq:final_result}, we calculate the distances moved by retinal reflex and the retinoscopic beam between a given time interval. However, since the smartphone camera is moving across frames, we need a static object as the point of reference for calculating the distance moved. We use the center of the fiducial markers as the origin since its position does not change in the video.

Next, we want to select the timestamps $t_{1}$ and $t_{2}$ between which to calculate the distance of retinal reflex and beam. We aim to choose frames with reflex edges clearly visible and which are farthest apart in time. Choosing frames farthest apart in time gives more robust estimates, as any localization errors in reflex / beam edge become smaller \textit{relative} to the overall motion. Note, however that since our formulation assumes standard optics assumptions (perfect lens, small angles, etc.), we want to stay near the center of the pupil. Thus, we restrict our search space to only the central 50\% of the pupil. We proceed by fitting a linear curve~\cite{Huber} on the position of reflex with respect to time. Then we select the start timestamp $t_{1}$ closest to the frame where the left/right edge of the retinal reflex covers at least 25\% of the pupil diameter, and the end timestamp $t_{2}$ closest to the frame where the same edge covers just beyond 75\% of the pupil diameter.

Next, to calculate the distance travelled by a reflex edge between the selected frames, we draw a line along the scoped meridian and measure the movement along this line. For robustness, we do this computation along multiple such lines and take the median distance. We do a similar computation for the retinoscopic beam. Finally, this is substituted into equation~\ref{eq:final_result} to obtain the refractive error.

%% file: sections/7_results.tex
\section{Results} \label{sec:results}
To evaluate the effectiveness of the proposed video analysis pipeline and the mathematical model, we conducted a clinical study where we collected retinoscopy videos at a local eye hospital. The retinoscopy videos were captured with the vertical streak beam scoping the horizontal meridian of the eye. Our video analysis pipeline predicts the net refractive error along the horizontal meridian. We compare our estimation with refractive error measurements from autorefractor (Topcon KR 8900), retinoscopy, and subjective refraction along the same horizontal meridian (using \textit{net refractive error = spherical + (cylindrical $\times$ $sin(axis)^2$)}, where $axis\in \{0^{\circ}-179^{\circ}\}$ \cite{meridional_refractometry}). We first examine the efficacy of the proposed video processing pipeline as a screening tool by classifying if the person has a refractive error or not, followed by a 4-class classification to understand the severity and type of refractive error. Next, we evaluate our estimated net refractive power with the ground truth subjective refraction measurements, and compare our results with prior work. 

\subsection{Clinical Study} \label{sec:clinicalResults}
The study included 185 distinct eyes with a mean refractive error of -1.13 $\pm$ 2.03D along horizontal meridian, ranging from -6.25D to 3.23D. A person should refer to an eye doctor for further tests if their refractive power is above 1D
or below -1D~\cite{myopia_cutoff}. We use the same cutoff values with the net refractive power for the binary classification---if the patient has a refractive error or not. Among the 185 eyes, 96 eyes were rated as normal (net refractive error between -1D and 1D), while the remaining 89 eyes had a refractive error, according to the subjective refraction measurements. Our proposed method achieves a sensitivity of 91.0\% and specificity of 74.0\% for this binary classification. Since we had an approximately uniform distribution for normal and refractive error cases, the high sensitivity value indicates that the proposed solution effectively recognizes patients with refractive power. However, our solution achieves a slightly low specificity of 74.0 \%, 
meaning normal eyes were marked as eyes with refractive error. In a medical setting, it is more critical not to let refractive error undiagnosed, hence our proposed solution has the potential to be used as a screening tool.

\input{tables/refractive_error_screening}

\begin{figure*}
\begin{center}
    \centering
    \includegraphics[width=1\linewidth]{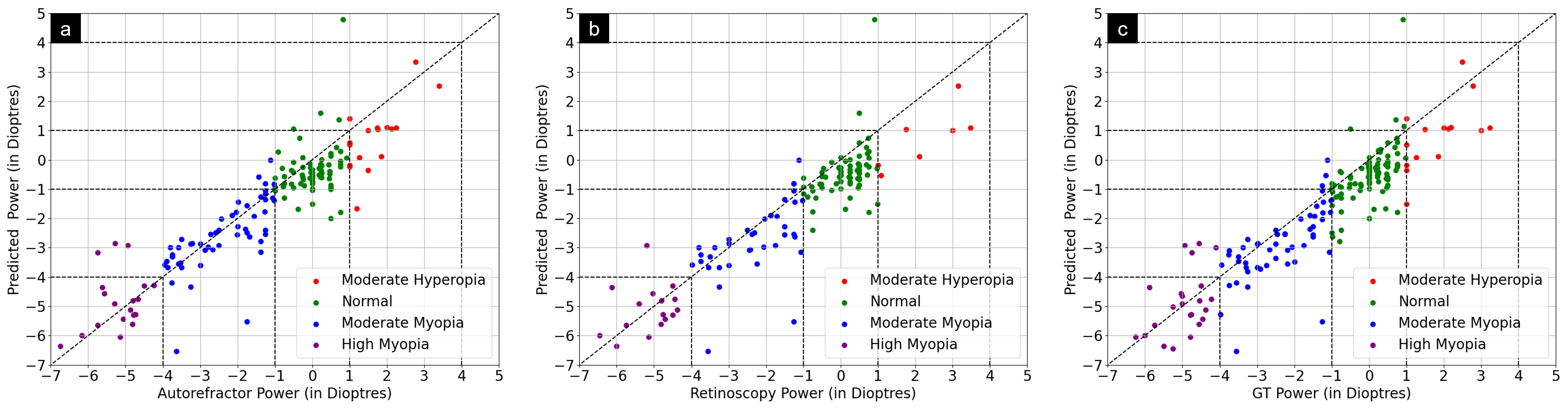}
\end{center}
    \caption{Correlation plots for refracted power predicted by our proposed system compared to (a) autorefractor, (b) manual retinoscopy, and (c) gold-standard subjective refraction.}
    \label{fig:mae_plot}
\end{figure*}

Next, we examine the type (myopia or hyperopia) and severity (moderate or high) of refractive error. For this we categorize our dataset into five classes: 
(a) high hyperopia (>= 4D), 
(b) moderate hyperopia (1D to 4D), 
(c) normal (-1D to 1D), 
(d) moderate myopia (-4D to -1D), and
(e) high myopia (<= -4D).
Due to the low prevalence of high hyperopia cases in the general population (<1\%, Figure~\ref{fig:distribution}a) and the small sample (3 eyes) in our dataset, we ignore them for this classification. For evaluating the 4-classes, we use one-vs-all approach, i.e., if the predicted class matches the ground truth class, we classify it as true positive, otherwise we consider it to be a false prediction. We achieve high sensitivity and specificity (>= 80 \%) for myopic eyes (Table~\ref{tab:classification}). This further supports our claim for the proposed system to be used as a screening tool because of the high prevalence of myopic cases in the general population (Figure~\ref{fig:distribution}a). For moderate hyperopia cases, we have a small sample size (of 15 eyes), and our approach achieves a low sensitivity of 60.0\%, which may be 
due to the steep gradient of ratio of distances moved by reflex and beam between two timestamps for +1D to +4D refractive power (Figure~\ref{fig:mathematical_model}). Existing approaches based on subjective refraction also report low sensitivity (69.2\%) and specificity (58.1\%) 
with hyperopic eyes, as evident in the prior work with 4497 eyes~\cite{subjective_hyperopia_challenge}.

\begin{figure*}
\begin{center}
    \centering
    \includegraphics[width=1\linewidth]{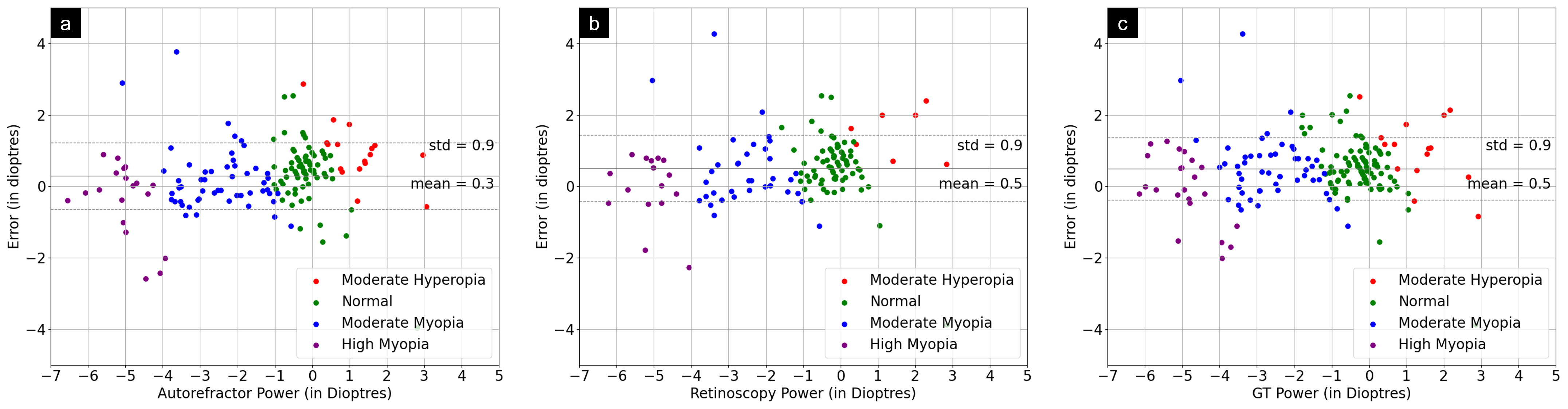}
\end{center}
    \caption{Bland-Altman plots for refracted power predicted by our proposed system compared to (a) autorefractor, (b) manual retinoscopy, and (c) gold-standard subjective refraction.}
    \label{fig:bland_plot}
\end{figure*}

\input{tables/mae_results}

We compare our estimated net refractive error (along the horizontal meridian) against the subjective refraction (ground truth), manual retinoscopy, and autorefractor measurements collected during the data collection (Table~\ref{tab:mae}). Among the 185 eyes in our dataset, we have autorefractor measurements for 161 eyes, and manual retinoscopy measurements for 130 eyes. Figure~\ref{fig:mae_plot}c shows a strong correlation between our predictions and the ground truth subjective refraction measurements, with the mean absolute error of 0.75 $\pm$ 0.67D. Similarly, Bland-Altman plots (Figure~\ref{fig:bland_plot}c) of the same measurements show a good agreement with a mean difference of 0.5 $\pm$ 0.9D.
The Bland-Altman plot also reveals the limits of agreement (95\%) as -1.26D to 2.26D. We also achieve a strong Pearson correlation coefficient, r(185) = 0.90, p < 0.01.
Figure~\ref{fig:mae_plot}a,b and Figure~\ref{fig:bland_plot}a,b show similar results, comparing our estimation with measurements from autorefractor and manual retinoscopy. Interestingly, 43.2\% of our predictions were within $\pm$0.5D, and 74.0\% were within $\pm$1D of the prescription given by an experienced optometrist using subjective refraction. These results are at par with the self-subjective refraction methods, like Eye Netra~\cite{netra_clinical} (38.0\% predictions within 0.5D) and USee~\cite{usee} (77.5\% predictions within 1D). Though these results are promising,
the mean absolute error of manual retinoscopy (0.18 $\pm$ 0.27D) and autorefractor (0.33 $\pm$ 0.33D) highlight a scope for improvement (refer to Table~\ref{tab:mae}).

\input{tables/other_devices_updated}
Furthermore, we compare our evaluation dataset statistics and performance of our retinoscopy-based approach with prior work (Table~\ref{tab:other_devices_updated}). Subjective refraction is used as ground truth measurement for reporting mean absolute error in all the methods.
Although smartphone-based approaches using eccentric photorefraction~\cite{smartphone_1_photorefraction, smartphone_2_photorefraction} report similar performance to our retinoscopy-based approach, their evaluation is limited to the pediatric population, mostly with myopic eyes.
On the flip side, self-subjective refraction-based methods (like Eye Netra~\cite{netra_clinical} and USee~\cite{usee}) evaluate only with the adult population highlighting their reliance on patient's cooperation. Meanwhile, our proposed automatic retinoscopy-based approach achieves comparable result on a diverse range of refractive power and age groups, in a low-cost manner. Expensive devices based on wavefront aberrometry (like QuickSee~\cite{Quicksee_AR_evaluation} and SVOne AR~\cite{sv_one_AR_clinic}) demonstrate the strongest agreement (< 0.5D) with the ground truth subjective refraction.

\subsection{Mathematical Model Analysis}
\label{sec:mathematical model results}
\begin{figure*}
\begin{center}
    \centering
    \includegraphics[width=0.5\linewidth]{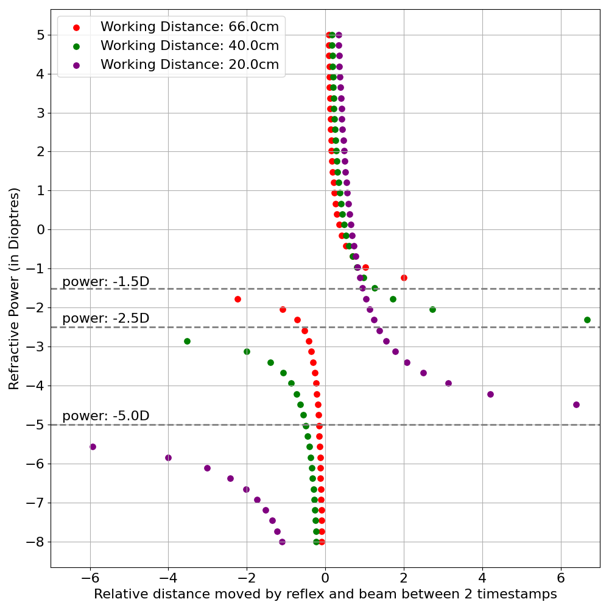}
\end{center}
    \caption{Plot of refractive power vs ratio of distance moved by retinal reflex and beam, between two timestamps, for three different working distances: 20cm, 40cm, and 66cm. Note: the optimal operating range decreases as the working distance increases.}
    \label{fig:mathematical_model}
\end{figure*}

Here we examine results from the mathematical model of retinoscopy (Section~\ref{sec:derivation}) to understand the effects of data collection parameters (working distance $d$ and effective source distance $u$). 
As the working distance decreases from 66 cm (used by optometrist for manual retinoscopy) to 40-30cm (used by our pipeline) with fixed $u$, the variation of $\Delta y_{p} / \Delta x_{p}$ increases for a fixed range of refractive error (Figure~\ref{fig:mathematical_model} plots the curve representing equation~\ref{eq:final_result}), thus enabling higher precision for the video analysis pipeline. Decreasing the working distance below 30cm should theoretically improve our results (Figure~\ref{fig:mathematical_model}). However, there are two reasons why we decided against a working distance of 20cm: (1) the patient starts accommodating as their view of the logMAR chart gets obstructed, and (2) the patient's pupil size decreases due to increased beam intensity from retinoscope on the patient's pupil. Hence we strike a trade-off at $\sim$30-40cm. 
For the effective source distance $u$, we selected the lowest vergence sleeve position as it was easiest to keep it consistent across the participants. For any other vergence sleeve position, an extra unnecessary step for calibrating $u$ would have been required.

%% file: tables/refractive_error_screening.tex
\begin{table}[!tb]
\centering
\caption{Sensitivity and Specificity for the 4-class refractive error screening}
\small
\begin{tabulary}{0.6\linewidth}{| c | c | c | c c |}
\toprule
\textbf{Class} & \textbf{Power Range} & \textbf{\# of eyes} & \textbf{Sensitivity} & \textbf{Specificity}\\ 
\hline
High Hyperopia & Above 4D & 3 & -- & --\\
Moderate Hyperopia & 1D to 4D & 15 & 60.0 \% & 98.2 \%\\
Normal & -1D to 1D & 96 & 74.0 \% & 91.0 \%\\
Moderate Myopia & -4D to -1D & 51 & 82.4 \% & 80.6 \%\\
High Myopia & Below -4D & 23 & 82.6 \% & 96.3 \%\\
 \bottomrule
\end{tabulary}
\label{tab:classification}
\end{table}

%% file: tables/mae_results.tex
\begin{table}[!tb]
\centering
\caption{Comparison of mean absolute error (MAE) $\pm$ standard deviation for the different measurements in our dataset.}
\small
\begin{tabulary}{0.6\linewidth}{| c | c | c | c |}
\toprule
\textbf{Method} & \textbf{Compared against} & \textbf{\# of eyes} & \textbf{MAE $\pm$ Std dev} \\ 
\hline
Manual Retinoscopy & Subjective Refraction & 130 & 0.18 $\pm$ 0.27D\\
Autorefractor & Subjective Refraction & 161 & 0.33 $\pm$ 0.33D\\
\hline
Ours & Manual Retinoscopy & 130 & 0.78 $\pm$ 0.72D\\
Ours & Autorefractor & 161 & 0.71 $\pm$ 0.67D\\
Ours & Subjective Refraction & 185 & 0.75 $\pm$ 0.67D\\
 \bottomrule
\end{tabulary}
\label{tab:mae}
\end{table}

%% file: tables/other_devices_updated.tex
\begin{table}[!tb]
\centering
\caption{Comparison of different approaches to estimate refractive error using portable devices. (*: Only myopic eyes)}
\small
\begin{tabulary}{0.95\linewidth}{| c | c | c | c | c | c |}
\toprule
\textbf{Principle} & \textbf{Method} & \textbf{\# of eyes}& \textbf{Mean Age} & \textbf{Power range} & \textbf{MAE $\pm$ Std dev} \\ 
\hline
\hline
{\multirow{2}{*}{Self-subjective Refraction}} & Eye Netra~\cite{netra_clinical} & 152 & 51.9 $\pm$ 18.3 years & -15.25D to 4.25D & 0.69 $\pm$ 0.53D\\ 
& USee~\cite{usee} & 120 & 53.1 $\pm$ 18.6 years & -6.00D to 5.00D & -\\
\hline
\hline
{\multirow{4}{*}{Eccentric Photorefraction}} & Plusoptix A09~\cite{plusoptix_clinical} & 64 & 58 months & - & 0.52 $\pm$ 1.20D\\
& SPOT~\cite{spot_clinical} & 134 & 29.7 years & - & 0.66 $\pm$ 0.56D\\ 
& \citet{smartphone_2_photorefraction} & 172 & 11.34 $\pm$ 0.96 years & -4.00D to 0.50D & 0.65D \\ 
& \citet{smartphone_1_photorefraction} & 165 & 10 years & <=0D & 0.80D \\
\hline
\hline
{\multirow{2}{*}{Wavefront Aberrometry}} & QuickSee~\cite{Quicksee_AR_evaluation} & 82 & 26.4 $\pm$ 9.7 years & -6.00D to 4.00D & 0.41 $\pm$ 0.53D \\
& SVOne AR*~\cite{sv_one_AR_clinic} & 92 & 24.8 years & -6.63D to 0D & 0.43D \\
\hline
\hline
Retinoscopy & Ours & 185 & 30.2 $\pm$ 11.1 years & -6.25D to 3.23D & 0.75 $\pm$ 0.67D\\
\bottomrule
\end{tabulary}
\label{tab:other_devices_updated}
\end{table}

%% file: sections/8_discussion.tex
\section{Discussion} \label{sec:discussion}
In this paper, our goal is to automate the process of retinoscopy for refractive error estimation by measuring the retinal reflex and retinoscopic beam movements. By doing so, we reduce the knowledge barrier to perform retinoscopy, eliminate the dependency on the expensive lens kit, and provide a means to preserve a digital record of retinoscopic videos for referring to it in the future. Our video processing pipeline tracks the retinoscopic beam and retinal reflex in the captured video and outputs net refractive error of the eye based on our extended mathematical formulation of retinoscopy. In our evaluation, we found that the proposed pipeline achieved a sensitivity and specificity of 91.0\% and 74.0\%, respectively, in diagnosing a patient's eye with a refractive error above 1D or below -1D. We also achieved an acceptable MAE (0.75 $\pm$ 0.67D) on a range of refractive powers as well as across different age groups. In comparison, prior portable devices for refractive error estimation are either very expensive, require a minimum skill level to operate or have limited evaluation. For example, low-cost devices based on eccentric photorefraction were evaluated on mainly pediatric population, while devices based on self-subjective refraction were evaluated on only adult population (Table~\ref{tab:other_devices_updated}).
Given the versatility of retinoscopy, we believe that capturing retinoscopic videos using a smartphone has the potential to be an important tool in a variety of eye screenings, and our work is the first step toward this goal. We discuss a few such applications.
Moreover, though our evaluation is limited to estimating net refractive error along the horizontal meridian, we discuss how this work can be extended to estimate spherical error, cylindrical error and its axis.

\subsection{Applications}
Uncorrected refractive error is a widespread vision problem, which can lead to permanent vision impairment if not addressed timely. Hence many portable medical-grade autorefractors are available to democratize eye care services. However, these devices are still expensive for the Global South, and they only serve a single purpose. 
In contrast, the videos that we captured during our retinoscopy procedure can be used for multiple applications, although we focused on estimating refractive error in this paper.

\subsubsection{Amblyopia Screening}
Undiagnosed refractive error can also cause amblyopia, a serious eye disorder. Amblyopia, also called lazy eye, mainly occurs due to poor vision in one eye, which can cause the brain to ignore the blurred image coming from the weaker eye. Over time this can lead to permanent vision loss in the weaker eye. 
Amblyopia affects around 3 out of every 100 children~\cite{kanna}. If diagnosed early (children of age 7 years or less), it is treatable by eye patching and wearing corrective lenses.

According to the AAPOS 2013 guidelines (for age >= 4 years)~\cite{AAPOS_2013}, amblyopia risk factors are: (a) difference in refractive power between the two eyes (anisometropia) > 1.5D, (b) hyperopia > 3.5D (along any meridian), (c) myopia < -1.5D (along any meridian), (d) astigmatism > 1.5D, (e) crossed eyes, and (f) media opacity >1 mm in size.
The first four risk factors are dependent on refractive error estimation, which can be measured by our proposed pipeline. Moreover, our approach can be extended to detect crossed eyes from the retinoscopy videos, similar to~\citet{strabismus_detect}.
Hence, our proposed low-cost portable solution has the potential to be useful in remote and resource-constrained settings~\cite{bhat-tc-cscw21} for amblyopia screening.

\subsubsection{Keratoconus Detection}
Keratoconus is a severe eye disease that leads to deformation of the cornea, developing a conical bulge.
It affects people in the age group of 10-25 years and progresses slowly for several years. If diagnosed early, it can be treated with corrective lenses; however, in advanced stages, it requires a corneal transplant or
can lead to partial or complete blindness. 
Thus timely diagnosis and treatment are essential to cure keratoconus. The diagnosis of keratoconus is performed by expensive and bulky medical devices called corneal topographers, which are not accessible to the masses, especially to people living in low- and middle-income countries. 
In a recent work~\cite{smartkc, gairola2022kcn}, a low-cost smartphone-based corneal topographer, SmartKC, was proposed to diagnose keratoconus. However, it requires additional hardware like a 3D printed placido head, USB-powered LEDs, and a paper-based diffuser. Although it achieves high accuracy in keratoconus diagnosis, the additional hardware makes it difficult to assemble by non-experts and requires additional training to use effectively.

On the other hand, retinoscopy has been shown to be very sensitive and reliable for detecting keratoconus even in the early stages of the disease~\cite{retino_kerato_1, retino_kerato_2}.
In keratoconus eyes, the retinal reflex, instead of being evenly distributed and moving in a single direction, shows two edges that move towards each other. This creates the visual effect of a scissor and is known as the \textit{scissors reflex}. A simple extension to our video processing pipeline can be made to identify such scissors reflex accurately. With that, our proposed system has the potential to be used for mass screening of keratoconus.

\subsubsection{Generate Complete Eye Prescription}
For evaluating our current system, we computed the net refractive power along the horizontal meridian ($0^o$). However, our approach is not limited to a single meridian. We can perform the same retinoscopy process (\textit{i.e.}, collect retinoscopy videos from the smartphone camera) along different meridians to estimate the net refractive power along those meridians. Leveraging the formulation from meridional refractometry~\cite{meridional_refractometry}, given the net refractive power along three predetermined meridians of the eye, the complete prescription of the eye (comprising of spherical power, cylindrical power, and cylindrical axis) can be calculated.

\subsubsection{Training Device} 
Retinoscopy is dependent on the examiner's perception of neutrality and has been found to be prone to inter-observer variability~\cite{retinoscopy_precision, PMID:28030881}. Moreover, prior work showed a direct relation between accuracy of refractive error estimation using retinoscopy with years of experience~\cite{retinoscopy_precision_2}, thus highlighting the requirement of highly-trained examiners with a certain level of experience.
At the same time, learning to use a retinoscope device is known to be hard, e.g., American Association of Ophthalmology states ``\textit{Retinoscopy is not an easy skill to learn. It takes patience and a lot of practice.}''~\cite{aao-retinoscopy}.

Looking beyond identifying eye disorders, our system can act as a guide for teaching retinoscopy to medical students. The ability to digitally zoom, pause/play/replay a recorded video, and track progress over time, can be valuable to explain, learn, document, and demonstrate abnormal retinal reflex movements and retinal disorders. Moreover, it can help bridge the communication gap between a teacher and students. There has been prior work~\cite{digital_retino} supporting the use of digital retinoscope as a training device.

\subsection{Real-time Assistance}
\label{sec: real_time_assistance}
During the pilot stage of data collection, we iteratively experimented with different parameters such as working distance, room brightness and beam width. Despite fixing these data collection parameters, we had to discard $\sim$27\% of collected videos due to small pupil size, noisy reflex/fiducial, blurry eyes, and out-of-bound working distance. 
To make the setup more user-friendly, we plan to implement on-device image processing based quality checks to provide real-time feedback to the operator. First, working distance should be calculated on the preview screen to guide the operator to move forward or backward while collecting data. Second, 
computing the sharpness (similar to \cite{smartkc, rdtscan}) of the eye region and fiducial markers while capturing the video can prompt the operator to recapture video if either of the regions is out-of-focus/blurry. Third, another problem that we identified in our data was the speed at which the retinoscope was rotated by the data collector.
As our video processing pipeline tracks retinal reflex throughout the video, it requires a slow movement of the retinoscope. In future deployments, the smartphone app should provide feedback to the operator regarding the speed of movement of the retinoscopic beam based on the readings from the smartphone's in-built IMU sensors.
Finally, thresholds like small pupil size (below 3mm) can be computed in real-time. In such a case, the app can suggest that the operator use cycloplegia drops, to reduce accommodation and increase pupil size.

\subsection{Failure Cases}
\label{sec:failureCases}
\begin{figure*}
\begin{center}
    \centering
    \includegraphics[width=\linewidth]{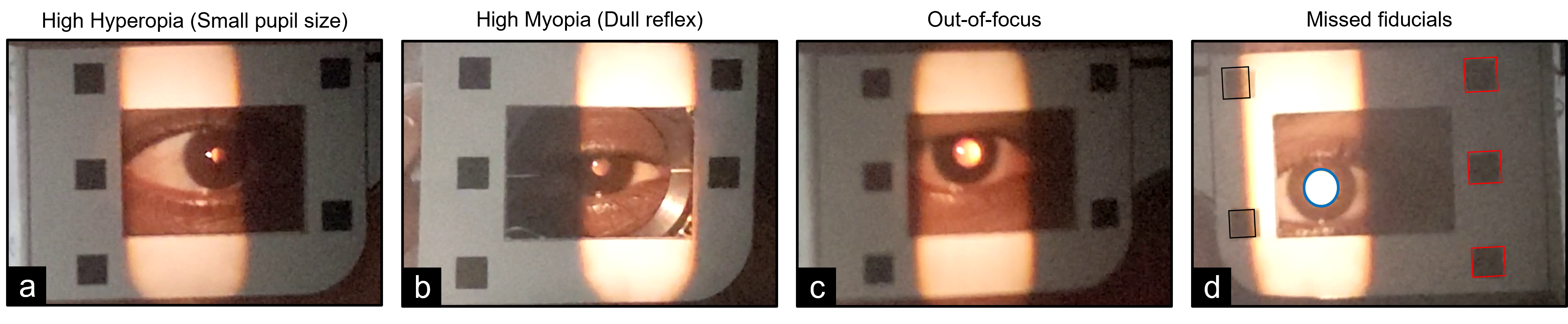}
\end{center}
    \caption{Cases where our proposed solution failed to estimate refractive error accurately: (a, b) outside operating range, and (c, d) data collection issues.
    }
    \label{fig:failure_cases}
\end{figure*}

Despite a robust and reliable video processing pipeline, we identified four types of cases (Figure~\ref{fig:failure_cases}) where our proposed solution failed to estimate refractive error correctly. These cases can be classified into two categories: outside operating range (Figure~\ref{fig:failure_cases}a,b) and data collection issues (Figure~\ref{fig:failure_cases}c,d).

First, in high hyperopia cases (> 4D), the pupil size of the patient is significantly smaller than the average pupil size~\cite{pupil_size} (Figure~\ref{fig:failure_cases}a). In such cases, 
our pipeline failed to track the retinal reflex movement within the pupil. 
Second, in high myopia cases (< -6D), the retinal reflex tends to be dull~\cite{corboy2003retinoscopy} (Figure~\ref{fig:failure_cases}b), again resulting in incorrect refractive error estimate. These problems are even faced by trained examiners while performing manual retinoscopy. However, since manual retinoscopy is a trial and error process, the retinal reflex becomes magnified and brighter as the examiner reaches neutrality by adding combinations of lenses in the trial frame. These two scenarios restrict our operating range from -6.0D to +3.0D, comprising \textasciitilde95\% of the clinical cases (Figure~\ref{fig:distribution}a).

Third, a few recorded videos were out-of-focus and/or noisy (Figure~\ref{fig:failure_cases}c). This resulted in blurred reflex edges, which were not correctly detected by our video processing pipeline. 
This can be identified and corrected with on-device image processing based quality check (similar to~\citet{smartkc}). Fourth, erroneous fiducial detection (Figure~\ref{fig:failure_cases}d) resulted in a significant deviation from the ground truth. This mainly happened in frames where the beam edges are close to the fiducial, resulting in interference. It can also be recognized and corrected during data collection by ensuring that the patient's pupil is approximately positioned at the center of the `reflex search space' of the paper frame (Figure~\ref{fig:glasses}c).

\subsection{Limitations}
We acknowledge a few limitations of our current work. First, due to COVID-19, we were able to recruit very few patients, hence
we were unable to collect a larger dataset (though it is comparable to recent mobile health sensing work~\cite{smartkc, seismo, BiliScreen}). 
Although our result seems encouraging on the limited range of refractive error, we believe an evaluation with a larger sample size across the range of refractive errors and age group is needed before real-world deployment. Second, our current evaluation is limited to estimating net refractive error along the horizontal meridian. Modern autorefractors output spherical power, cylindrical power, and cylindrical axis. Our setup can also achieve that by computing spherical power along three predetermined axes (using \cite{meridional_refractometry}), though the accuracy of that approach needs to be rigorously evaluated. Third, in the current setup we used a medical-grade retinoscope attached to a smartphone camera for data collection.
In the future, we plan to use a cheap retinoscope, or a 3D-printed attachment imitating a retinoscope (similar to \cite{smartkc, gairola2022kcn}), thus reducing the setup cost. Finally, for the current evaluation, the data was collected with a single smartphone by two data collectors. For wider deployment, we need to further evaluate our results on other smartphones with more data collectors.

%% file: sections/9_conclusion.tex
\section{Conclusion} \label{sec:conclusion}
Automating retinoscopy can be an important tool for the early diagnosis of refractive error, a leading cause of vision impairment globally. Along with all the advantages of a retinoscope (like low-cost, portable, and no cooperation required from the patient), a digital retinoscope offers two key advantages---no requirement of a lens kit and no requirement of a trained examiner.
In this work, we extended the mathematical formulation of retinoscopy to work on smartphone camera captured retinoscopic videos, and developed a video processing pipeline based on that formulation.
We collected 185 eyes video data by attaching a smartphone to a retinoscope with the patient wearing a paper frame. Our approach achieved a sensitivity of 91.0\% and specificity of 74.0\% for refractive error classification, and a mean absolute error of 0.75$\pm$0.67D for net refractive error estimation.
Although more research is needed to develop a functional smartphone-based digital retinoscope at par with commercial medical devices, we believe that this is the first step in showing the feasibility of such a device.

%% file: main.bbl
%%% -*-BibTeX-*-
%%% Do NOT edit. File created by BibTeX with style
%%% ACM-Reference-Format-Journals [18-Jan-2012].

\begin{thebibliography}{64}

%%% ====================================================================
%%% NOTE TO THE USER: you can override these defaults by providing
%%% customized versions of any of these macros before the \bibliography
%%% command.  Each of them MUST provide its own final punctuation,
%%% except for \shownote{}, \showDOI{}, and \showURL{}.  The latter two
%%% do not use final punctuation, in order to avoid confusing it with
%%% the Web address.
%%%
%%% To suppress output of a particular field, define its macro to expand
%%% to an empty string, or better, \unskip, like this:
%%%
%%% \newcommand{\showDOI}[1]{\unskip}   % LaTeX syntax
%%%
%%% \def \showDOI #1{\unskip}           % plain TeX syntax
%%%
%%% ====================================================================

\ifx \showCODEN    \undefined \def \showCODEN     #1{\unskip}     \fi
\ifx \showDOI      \undefined \def \showDOI       #1{#1}\fi
\ifx \showISBNx    \undefined \def \showISBNx     #1{\unskip}     \fi
\ifx \showISBNxiii \undefined \def \showISBNxiii  #1{\unskip}     \fi
\ifx \showISSN     \undefined \def \showISSN      #1{\unskip}     \fi
\ifx \showLCCN     \undefined \def \showLCCN      #1{\unskip}     \fi
\ifx \shownote     \undefined \def \shownote      #1{#1}          \fi
\ifx \showarticletitle \undefined \def \showarticletitle #1{#1}   \fi
\ifx \showURL      \undefined \def \showURL       {\relax}        \fi
% The following commands are used for tagged output and should be
% invisible to TeX
\providecommand\bibfield[2]{#2}
\providecommand\bibinfo[2]{#2}
\providecommand\natexlab[1]{#1}
\providecommand\showeprint[2][]{arXiv:#2}

\bibitem[Agarwal et~al\mbox{.}(2019)]%
        {netra_time}
\bibfield{author}{\bibinfo{person}{Arunika Agarwal}, \bibinfo{person}{David~E
  Bloom}, \bibinfo{person}{Vincent~P deLuise}, \bibinfo{person}{Alyssa Lubet},
  \bibinfo{person}{Kaushik Murali}, {and} \bibinfo{person}{Srinivas~M Sastry}.}
  \bibinfo{year}{2019}\natexlab{}.
\newblock \showarticletitle{Comparing low-cost handheld autorefractors: A
  practical approach to measuring refraction in low-resource settings}.
\newblock \bibinfo{journal}{\emph{PLoS One}} \bibinfo{volume}{14},
  \bibinfo{number}{10} (\bibinfo{date}{Oct} \bibinfo{year}{2019}).
\newblock


\bibitem[Al-Mahrouqi et~al\mbox{.}(2019)]%
        {retino_kerato_1}
\bibfield{author}{\bibinfo{person}{Haitham Al-Mahrouqi}, \bibinfo{person}{Saif
  Oraba}, \bibinfo{person}{Shihab Al-Habsi}, \bibinfo{person}{Noufal
  Mundemkattil}, \bibinfo{person}{Jithin Babu}, \bibinfo{person}{Sathiya
  Panchatcharam}, \bibinfo{person}{Rashid Al-Saidi}, {and}
  \bibinfo{person}{Abdulatif Al-Raisi}.} \bibinfo{year}{2019}\natexlab{}.
\newblock \showarticletitle{Retinoscopy as a Screening Tool for Keratoconus}.
\newblock \bibinfo{journal}{\emph{Cornea}}  \bibinfo{volume}{38}
  (\bibinfo{date}{04} \bibinfo{year}{2019}), \bibinfo{pages}{1}.
\newblock
\urldef\tempurl%
\url{https://doi.org/10.1097/ICO.0000000000001843}
\showDOI{\tempurl}


\bibitem[{American Optometric Association}(2020)]%
        {aao_comprehensive_pediatric_eye}
\bibfield{author}{\bibinfo{person}{{American Optometric Association}}.}
  \bibinfo{year}{2020}\natexlab{}.
\newblock \showarticletitle{Evidence-based clinical practice guideline:
  Comprehensive pediatric eye and vision examination}.
\newblock \bibinfo{journal}{\emph{Optometric Clinical Practice}}
  (\bibinfo{year}{2020}), \bibinfo{pages}{2--67}.
\newblock


\bibitem[Annadanam et~al\mbox{.}(2018)]%
        {usee}
\bibfield{author}{\bibinfo{person}{Anvesh Annadanam}, \bibinfo{person}{Varshini
  Varadaraj}, \bibinfo{person}{Lucy Mudie}, \bibinfo{person}{Alice Liu},
  \bibinfo{person}{William Plum}, \bibinfo{person}{J. White},
  \bibinfo{person}{Megan Collins}, {and} \bibinfo{person}{David Friedman}.}
  \bibinfo{year}{2018}\natexlab{}.
\newblock \showarticletitle{Comparison of self-refraction using a simple
  device, USee, with manifest refraction in adults}.
\newblock \bibinfo{journal}{\emph{PLOS ONE}}  \bibinfo{volume}{13}
  (\bibinfo{date}{02} \bibinfo{year}{2018}).
\newblock
\urldef\tempurl%
\url{https://doi.org/10.1371/journal.pone.0192055}
\showDOI{\tempurl}


\bibitem[Bharadwaj et~al\mbox{.}(2014)]%
        {retinoscopy_precision_2}
\bibfield{author}{\bibinfo{person}{Shrikant~R Bharadwaj},
  \bibinfo{person}{Menaka Malavita}, {and} \bibinfo{person}{Jennifer Jayaraj}.}
  \bibinfo{year}{2014}\natexlab{}.
\newblock \showarticletitle{A psychophysical technique for estimating the
  accuracy and precision of retinoscopy}.
\newblock \bibinfo{journal}{\emph{Clinical and Experimental Optometry}}
  \bibinfo{volume}{97}, \bibinfo{number}{2} (\bibinfo{year}{2014}),
  \bibinfo{pages}{164--170}.
\newblock
\urldef\tempurl%
\url{https://doi.org/10.1111/cxo.12112}
\showDOI{\tempurl}


\bibitem[Bhat et~al\mbox{.}(2021)]%
        {bhat-tc-cscw21}
\bibfield{author}{\bibinfo{person}{Karthik~S Bhat}, \bibinfo{person}{Mohit
  Jain}, {and} \bibinfo{person}{Neha Kumar}.} \bibinfo{year}{2021}\natexlab{}.
\newblock \showarticletitle{Infrastructuring Telehealth in (In)Formal
  Patient-Doctor Contexts}.
\newblock \bibinfo{journal}{\emph{Proc. ACM Hum.-Comput. Interact.}}
  \bibinfo{volume}{5}, \bibinfo{number}{CSCW2}, Article
  \bibinfo{articleno}{323} (\bibinfo{date}{oct} \bibinfo{year}{2021}),
  \bibinfo{numpages}{28}~pages.
\newblock
\urldef\tempurl%
\url{https://doi.org/10.1145/3476064}
\showDOI{\tempurl}


\bibitem[Boeder and Kolder(1984)]%
        {retino_derivation}
\bibfield{author}{\bibinfo{person}{Paul Boeder} {and} \bibinfo{person}{H.~E.
  Kolder}.} \bibinfo{year}{1984}\natexlab{}.
\newblock \showarticletitle{{Neutralization at Infinity in Streak
  Retinoscopy}}.
\newblock \bibinfo{journal}{\emph{Archives of Ophthalmology}}
  \bibinfo{volume}{102}, \bibinfo{number}{9} (\bibinfo{date}{09}
  \bibinfo{year}{1984}), \bibinfo{pages}{1396--1399}.
\newblock
\showISSN{0003-9950}
\urldef\tempurl%
\url{https://doi.org/10.1001/archopht.1984.01040031138042}
\showDOI{\tempurl}
\showeprint{https://jamanetwork.com/journals/jamaophthalmology/articlepdf/635237/archopht\_102\_9\_042.pdf}


\bibitem[Bourne et~al\mbox{.}(2020)]%
        {30_years_trends}
\bibfield{author}{\bibinfo{person}{Rupert Bourne}, \bibinfo{person}{Jaimie
  Adelson}, \bibinfo{person}{Seth Flaxman}, \bibinfo{person}{Paul Briant},
  \bibinfo{person}{Hugh Taylor}, \bibinfo{person}{Robert Casson},
  \bibinfo{person}{Mukharram Bikbov}, \bibinfo{person}{Michele Bottone},
  \bibinfo{person}{Tasanee Braithwaite}, \bibinfo{person}{Alain Bron},
  \bibinfo{person}{Ching-yu Cheng}, \bibinfo{person}{Maria~Vittoria Cicinelli},
  \bibinfo{person}{Nathan Congdon}, \bibinfo{person}{Arthur Fernandes},
  \bibinfo{person}{David Friedman}, \bibinfo{person}{Joao Furtado},
  \bibinfo{person}{Ronnie George}, \bibinfo{person}{Rim Kahloun},
  \bibinfo{person}{John Kempen}, {and} \bibinfo{person}{Theo Vos}.}
  \bibinfo{year}{2020}\natexlab{}.
\newblock \showarticletitle{Trends in Prevalence of Blindness and Distance and
  Near Vision Impairment Over 30 Years and Contribution to the Global Burden of
  Disease in 2020}.
\newblock \bibinfo{journal}{\emph{SSRN Electronic Journal}} (\bibinfo{date}{01}
  \bibinfo{year}{2020}).
\newblock
\urldef\tempurl%
\url{https://doi.org/10.2139/ssrn.3582742}
\showDOI{\tempurl}


\bibitem[Brubaker et~al\mbox{.}(1969)]%
        {meridional_refractometry}
\bibfield{author}{\bibinfo{person}{Richard~F. Brubaker},
  \bibinfo{person}{Robert~D. Reinecke}, {and} \bibinfo{person}{Jack~C.
  Copeland}.} \bibinfo{year}{1969}\natexlab{}.
\newblock \showarticletitle{{Meridional Refractometry: I. Derivation of
  Equations}}.
\newblock \bibinfo{journal}{\emph{Archives of Ophthalmology}}
  \bibinfo{volume}{81}, \bibinfo{number}{6} (\bibinfo{date}{06}
  \bibinfo{year}{1969}), \bibinfo{pages}{849--852}.
\newblock
\showISSN{0003-9950}
\urldef\tempurl%
\url{https://doi.org/10.1001/archopht.1969.00990010851018}
\showDOI{\tempurl}


\bibitem[Buades et~al\mbox{.}(2011)]%
        {ipol.2011.bcm_nlm}
\bibfield{author}{\bibinfo{person}{Antoni Buades}, \bibinfo{person}{Bartomeu
  Coll}, {and} \bibinfo{person}{Jean-Michel Morel}.}
  \bibinfo{year}{2011}\natexlab{}.
\newblock \showarticletitle{{Non-Local Means Denoising}}.
\newblock \bibinfo{journal}{\emph{{Image Processing On Line}}}
  \bibinfo{volume}{1} (\bibinfo{year}{2011}), \bibinfo{pages}{208--212}.
\newblock
\newblock
\shownote{\url{https://doi.org/10.5201/ipol.2011.bcm_nlm}}.


\bibitem[Canny(1986)]%
        {Canny}
\bibfield{author}{\bibinfo{person}{John Canny}.}
  \bibinfo{year}{1986}\natexlab{}.
\newblock \showarticletitle{A Computational Approach to Edge Detection}.
\newblock \bibinfo{journal}{\emph{IEEE Transactions on Pattern Analysis and
  Machine Intelligence}} \bibinfo{volume}{PAMI-8}, \bibinfo{number}{6}
  (\bibinfo{year}{1986}), \bibinfo{pages}{679--698}.
\newblock
\urldef\tempurl%
\url{https://doi.org/10.1109/TPAMI.1986.4767851}
\showDOI{\tempurl}


\bibitem[Chan et~al\mbox{.}(2014)]%
        {digital_retino}
\bibfield{author}{\bibinfo{person}{Weng~Onn Chan}, \bibinfo{person}{Matthew
  Crabb}, \bibinfo{person}{David Sia}, {and} \bibinfo{person}{Deepa~Ajay
  Taranath}.} \bibinfo{year}{2014}\natexlab{}.
\newblock \showarticletitle{Creating a digital retinoscope by combining a
  mobile smartphone camera and a retinoscope.}
\newblock \bibinfo{journal}{\emph{Journal of AAPOS : the official publication
  of the American Association for Pediatric Ophthalmology and Strabismus}}
  \bibinfo{volume}{18 4} (\bibinfo{year}{2014}), \bibinfo{pages}{387--8}.
\newblock


\bibitem[Chen et~al\mbox{.}(2003)]%
        {eccentric_simulation}
\bibfield{author}{\bibinfo{person}{Ying-Ling Chen}, \bibinfo{person}{bo Tan},
  {and} \bibinfo{person}{Jwl Lewis}.} \bibinfo{year}{2003}\natexlab{}.
\newblock \showarticletitle{Simulation of eccentric photorefraction images}.
\newblock \bibinfo{journal}{\emph{Optics express}}  \bibinfo{volume}{11}
  (\bibinfo{date}{08} \bibinfo{year}{2003}), \bibinfo{pages}{1628--42}.
\newblock
\urldef\tempurl%
\url{https://doi.org/10.1364/OE.11.001628}
\showDOI{\tempurl}


\bibitem[Choong et~al\mbox{.}(2006)]%
        {Choong2006-vy}
\bibfield{author}{\bibinfo{person}{Yee-Fong Choong}, \bibinfo{person}{Ai-Hong
  Chen}, {and} \bibinfo{person}{Pik-Pin Goh}.} \bibinfo{year}{2006}\natexlab{}.
\newblock \showarticletitle{A comparison of autorefraction and subjective
  refraction with and without cycloplegia in primary school children}.
\newblock \bibinfo{journal}{\emph{Am J Ophthalmol}} \bibinfo{volume}{142},
  \bibinfo{number}{1} (\bibinfo{date}{July} \bibinfo{year}{2006}),
  \bibinfo{pages}{68--74}.
\newblock


\bibitem[Ciuffreda and Rosenfield(2015)]%
        {sv_one_AR_clinic}
\bibfield{author}{\bibinfo{person}{Kenneth Ciuffreda} {and}
  \bibinfo{person}{Mark Rosenfield}.} \bibinfo{year}{2015}\natexlab{}.
\newblock \showarticletitle{Evaluation of the SVOne: A Handheld,
  Smartphone-Based Autorefractor}.
\newblock \bibinfo{journal}{\emph{Optometry and vision science : official
  publication of the American Academy of Optometry}}  \bibinfo{volume}{92}
  (\bibinfo{date}{11} \bibinfo{year}{2015}).
\newblock
\urldef\tempurl%
\url{https://doi.org/10.1097/OPX.0000000000000726}
\showDOI{\tempurl}


\bibitem[Corboy(2003)]%
        {corboy2003retinoscopy}
\bibfield{author}{\bibinfo{person}{J.M. Corboy}.}
  \bibinfo{year}{2003}\natexlab{}.
\newblock \bibinfo{booktitle}{\emph{The Retinoscopy Book: An Introductory
  Manual for Eye Care Professionals}}.
\newblock \bibinfo{publisher}{Slack}.
\newblock
\showISBNx{9781556426230}
\showLCCN{2003002379}
\urldef\tempurl%
\url{https://books.google.co.in/books?id=6I6JeDWonhQC}
\showURL{%
\tempurl}


\bibitem[Cosme-Cisneros et~al\mbox{.}(2017)]%
        {Purkinje}
\bibfield{author}{\bibinfo{person}{I. Cosme-Cisneros}, \bibinfo{person}{G.
  Escamilla-Ruiz}, \bibinfo{person}{D. Flores-Montoya}, \bibinfo{person}{G.
  Hernández-Gómez}, {and} \bibinfo{person}{A. Gómez-Vieyra}.}
  \bibinfo{year}{2017}\natexlab{}.
\newblock \bibinfo{booktitle}{\emph{Instrument for Recording Purkinje Images}}.
\newblock \bibinfo{pages}{101--109}.
\newblock
\showISBNx{978-3-319-28511-5}
\urldef\tempurl%
\url{https://doi.org/10.1007/978-3-319-28513-9_14}
\showDOI{\tempurl}


\bibitem[Dandona et~al\mbox{.}(2001)]%
        {india-referror}
\bibfield{author}{\bibinfo{person}{L Dandona}, \bibinfo{person}{R Dandona},
  \bibinfo{person}{M Srinivas}, \bibinfo{person}{P Giridhar},
  \bibinfo{person}{K Vilas}, \bibinfo{person}{M~N Prasad}, \bibinfo{person}{R~K
  John}, \bibinfo{person}{C~A McCarty}, {and} \bibinfo{person}{G~N Rao}.}
  \bibinfo{year}{2001}\natexlab{}.
\newblock \showarticletitle{Blindness in the Indian state of Andhra Pradesh}.
\newblock \bibinfo{journal}{\emph{Invest Ophthalmol Vis Sci}}
  \bibinfo{volume}{42}, \bibinfo{number}{5} (\bibinfo{date}{April}
  \bibinfo{year}{2001}), \bibinfo{pages}{908--916}.
\newblock


\bibitem[Donahue et~al\mbox{.}(2013)]%
        {AAPOS_2013}
\bibfield{author}{\bibinfo{person}{Sean~P. Donahue}, \bibinfo{person}{Brian
  Arthur}, \bibinfo{person}{Daniel~E. Neely}, \bibinfo{person}{Robert~W.
  Arnold}, \bibinfo{person}{David Silbert}, {and} \bibinfo{person}{James~B.
  Ruben}.} \bibinfo{year}{2013}\natexlab{}.
\newblock \showarticletitle{Guidelines for automated preschool vision
  screening: A 10-year, evidence-based update}.
\newblock \bibinfo{journal}{\emph{Journal of American Association for Pediatric
  Ophthalmology and Strabismus}} \bibinfo{volume}{17}, \bibinfo{number}{1}
  (\bibinfo{year}{2013}), \bibinfo{pages}{4--8}.
\newblock
\showISSN{1091-8531}
\urldef\tempurl%
\url{https://doi.org/10.1016/j.jaapos.2012.09.012}
\showDOI{\tempurl}


\bibitem[Dong et~al\mbox{.}(2016)]%
        {FSRCNN}
\bibfield{author}{\bibinfo{person}{Chao Dong}, \bibinfo{person}{Chen~Change
  Loy}, {and} \bibinfo{person}{Xiaoou Tang}.} \bibinfo{year}{2016}\natexlab{}.
\newblock \showarticletitle{Accelerating the Super-Resolution Convolutional
  Neural Network}. In \bibinfo{booktitle}{\emph{Computer Vision -- ECCV 2016}},
  \bibfield{editor}{\bibinfo{person}{Bastian Leibe}, \bibinfo{person}{Jiri
  Matas}, \bibinfo{person}{Nicu Sebe}, {and} \bibinfo{person}{Max Welling}}
  (Eds.). \bibinfo{publisher}{Springer International Publishing},
  \bibinfo{address}{Cham}, \bibinfo{pages}{391--407}.
\newblock
\showISBNx{978-3-319-46475-6}


\bibitem[Durr et~al\mbox{.}(2015)]%
        {Quicksee_AR_evaluation}
\bibfield{author}{\bibinfo{person}{Nicholas~J Durr}, \bibinfo{person}{Shivang~R
  Dave}, \bibinfo{person}{Fuensanta~A Vera-Diaz}, \bibinfo{person}{Daryl Lim},
  \bibinfo{person}{Carlos Dorronsoro}, \bibinfo{person}{Susana Marcos},
  \bibinfo{person}{Frank Thorn}, {and} \bibinfo{person}{Eduardo Lage}.}
  \bibinfo{year}{2015}\natexlab{}.
\newblock \showarticletitle{Design and Clinical Evaluation of a Handheld
  Wavefront Autorefractor}.
\newblock \bibinfo{journal}{\emph{Optometry and vision science : official
  publication of the American Academy of Optometry}} \bibinfo{volume}{92},
  \bibinfo{number}{12} (\bibinfo{date}{December} \bibinfo{year}{2015}),
  \bibinfo{pages}{1140—1147}.
\newblock
\showISSN{1040-5488}
\urldef\tempurl%
\url{https://doi.org/10.1097/opx.0000000000000732}
\showDOI{\tempurl}


\bibitem[Elliott(2016)]%
        {logmarChart}
\bibfield{author}{\bibinfo{person}{David~B. Elliott}.}
  \bibinfo{year}{2016}\natexlab{}.
\newblock \showarticletitle{The good (logMAR), the bad (Snellen) and the ugly
  (BCVA, number of letters read) of visual acuity measurement}.
\newblock \bibinfo{journal}{\emph{Ophthalmic and Physiological Optics}}
  \bibinfo{volume}{36}, \bibinfo{number}{4} (\bibinfo{year}{2016}),
  \bibinfo{pages}{355--358}.
\newblock
\urldef\tempurl%
\url{https://doi.org/10.1111/opo.12310}
\showDOI{\tempurl}
\showeprint{https://onlinelibrary.wiley.com/doi/pdf/10.1111/opo.12310}


\bibitem[Flitcroft et~al\mbox{.}(2019)]%
        {myopia_cutoff}
\bibfield{author}{\bibinfo{person}{Daniel~Ian Flitcroft},
  \bibinfo{person}{Mingguang He}, \bibinfo{person}{Jost~B. Jonas},
  \bibinfo{person}{Monica Jong}, \bibinfo{person}{Kovin Naidoo},
  \bibinfo{person}{Kyoko Ohno-Matsui}, \bibinfo{person}{Jugnoo Rahi},
  \bibinfo{person}{Serge Resnikoff}, \bibinfo{person}{Susan Vitale}, {and}
  \bibinfo{person}{Lawrence Yannuzzi}.} \bibinfo{year}{2019}\natexlab{}.
\newblock \showarticletitle{{IMI – Defining and Classifying Myopia: A
  Proposed Set of Standards for Clinical and Epidemiologic Studies}}.
\newblock \bibinfo{journal}{\emph{Investigative Ophthalmology \& Visual
  Science}} \bibinfo{volume}{60}, \bibinfo{number}{3} (\bibinfo{date}{02}
  \bibinfo{year}{2019}), \bibinfo{pages}{M20--M30}.
\newblock
\showISSN{1552-5783}
\urldef\tempurl%
\url{https://doi.org/10.1167/iovs.18-25957}
\showDOI{\tempurl}
\showeprint{https://arvojournals.org/arvo/content\_public/journal/iovs/937872/i1552-5783-60-3-m20.pdf}


\bibitem[for Innovation(2018)]%
        {fofo}
\bibfield{author}{\bibinfo{person}{LVPEI~Center for Innovation}.}
  \bibinfo{year}{2018}\natexlab{}.
\newblock \bibinfo{booktitle}{\emph{FoFo (Folding Foropter) project}}.
\newblock
\urldef\tempurl%
\url{https://www.fofo-project.com/}
\showURL{%
Retrieved January 1,2022 from \tempurl}


\bibitem[Fu et~al\mbox{.}(2020)]%
        {smartphone_2_photorefraction}
\bibfield{author}{\bibinfo{person}{Eugene Fu}, \bibinfo{person}{Zhongqi Yang},
  \bibinfo{person}{Hong Leong}, \bibinfo{person}{Grace Ngai},
  \bibinfo{person}{Chi~Wai Do}, {and} \bibinfo{person}{Lily Chan}.}
  \bibinfo{year}{2020}\natexlab{}.
\newblock \showarticletitle{Exploiting Active Learning in Novel Refractive
  Error Detection with Smartphones}. \bibinfo{pages}{2775--2783}.
\newblock
\urldef\tempurl%
\url{https://doi.org/10.1145/3394171.3413748}
\showDOI{\tempurl}


\bibitem[Gairola et~al\mbox{.}(2022a)]%
        {smartkc}
\bibfield{author}{\bibinfo{person}{Siddhartha Gairola},
  \bibinfo{person}{Murtuza Bohra}, \bibinfo{person}{Nadeem Shaheer},
  \bibinfo{person}{Navya Jayaprakash}, \bibinfo{person}{Pallavi Joshi},
  \bibinfo{person}{Anand Balasubramaniam}, \bibinfo{person}{Kaushik Murali},
  \bibinfo{person}{Nipun Kwatra}, {and} \bibinfo{person}{Mohit Jain}.}
  \bibinfo{year}{2022}\natexlab{a}.
\newblock \showarticletitle{SmartKC: Smartphone-Based Corneal Topographer for
  Keratoconus Detection}.
\newblock \bibinfo{journal}{\emph{Proc. ACM Interact. Mob. Wearable Ubiquitous
  Technol.}} \bibinfo{volume}{5}, \bibinfo{number}{4}, Article
  \bibinfo{articleno}{155} (\bibinfo{date}{dec} \bibinfo{year}{2022}),
  \bibinfo{numpages}{27}~pages.
\newblock
\urldef\tempurl%
\url{https://doi.org/10.1145/3494982}
\showDOI{\tempurl}


\bibitem[Gairola et~al\mbox{.}(2022b)]%
        {gairola2022kcn}
\bibfield{author}{\bibinfo{person}{Siddhartha Gairola},
  \bibinfo{person}{Pallavi Joshi}, \bibinfo{person}{Anand Balasubramaniam},
  \bibinfo{person}{Kaushik Murali}, \bibinfo{person}{Nipun Kwatra}, {and}
  \bibinfo{person}{Mohit Jain}.} \bibinfo{year}{2022}\natexlab{b}.
\newblock \showarticletitle{Keratoconus Classifier for Smartphone-based Corneal
  Topographer}. In \bibinfo{booktitle}{\emph{2022 44th Annual International
  Conference of the IEEE Engineering in Medicine \& Biology Society (EMBC)}}.
  \bibinfo{publisher}{IEEE}.
\newblock


\bibitem[Gencer et~al\mbox{.}(2015)]%
        {pupil_size}
\bibfield{author}{\bibinfo{person}{Baran Gencer}, \bibinfo{person}{Engin
  Ozgurhan}, \bibinfo{person}{Tuğba Kurt}, \bibinfo{person}{Ismail Ersan},
  \bibinfo{person}{Sedat Arıkan}, \bibinfo{person}{Selçuk Kara},
  \bibinfo{person}{Mediha Coşar}, {and} \bibinfo{person}{Ercüment Bozkurt}.}
  \bibinfo{year}{2015}\natexlab{}.
\newblock \showarticletitle{Comparision of pupil size in myopic and hyperopic
  patients}.
\newblock \bibinfo{journal}{\emph{Pamukkale Medical Journal}}
  \bibinfo{volume}{8} (\bibinfo{date}{01} \bibinfo{year}{2015}),
  \bibinfo{pages}{88--91}.
\newblock
\urldef\tempurl%
\url{https://doi.org/10.5505/ptd.2015.59455}
\showDOI{\tempurl}


\bibitem[Hartley and Zisserman(2004)]%
        {Hartley2004}
\bibfield{author}{\bibinfo{person}{R.~I. Hartley} {and} \bibinfo{person}{A.
  Zisserman}.} \bibinfo{year}{2004}\natexlab{}.
\newblock \bibinfo{booktitle}{\emph{Multiple View Geometry in Computer Vision}
  (\bibinfo{edition}{second} ed.)}.
\newblock \bibinfo{publisher}{Cambridge University Press, ISBN: 0521540518}.
\newblock


\bibitem[Huang et~al\mbox{.}(2021)]%
        {strabismus_detect}
\bibfield{author}{\bibinfo{person}{Xilang Huang}, \bibinfo{person}{Sang~Joon
  Lee}, \bibinfo{person}{Chang Kim}, {and} \bibinfo{person}{Seon~Han Choi}.}
  \bibinfo{year}{2021}\natexlab{}.
\newblock \showarticletitle{An automatic screening method for strabismus
  detection based on image processing}.
\newblock \bibinfo{journal}{\emph{PLOS ONE}}  \bibinfo{volume}{16}
  (\bibinfo{date}{08} \bibinfo{year}{2021}), \bibinfo{pages}{e0255643}.
\newblock
\urldef\tempurl%
\url{https://doi.org/10.1371/journal.pone.0255643}
\showDOI{\tempurl}


\bibitem[Illingworth and Kittler(1987)]%
        {HOUGH_1}
\bibfield{author}{\bibinfo{person}{J. Illingworth} {and} \bibinfo{person}{J.
  Kittler}.} \bibinfo{year}{1987}\natexlab{}.
\newblock \showarticletitle{The Adaptive Hough Transform}.
\newblock \bibinfo{journal}{\emph{IEEE Transactions on Pattern Analysis and
  Machine Intelligence}} \bibinfo{volume}{PAMI-9}, \bibinfo{number}{5}
  (\bibinfo{year}{1987}), \bibinfo{pages}{690--698}.
\newblock
\urldef\tempurl%
\url{https://doi.org/10.1109/TPAMI.1987.4767964}
\showDOI{\tempurl}


\bibitem[Jeganathan et~al\mbox{.}(2018)]%
        {netra_clinical}
\bibfield{author}{\bibinfo{person}{V~Swetha~E Jeganathan},
  \bibinfo{person}{Nita Valikodath}, \bibinfo{person}{Leslie~M Niziol},
  \bibinfo{person}{Sean Hansen}, \bibinfo{person}{Hannah Apostolou}, {and}
  \bibinfo{person}{Maria~A Woodward}.} \bibinfo{year}{2018}\natexlab{}.
\newblock \showarticletitle{Accuracy of a Smartphone-based Autorefractor
  Compared with Criterion-standard Refraction}.
\newblock \bibinfo{journal}{\emph{Optometry and vision science : official
  publication of the American Academy of Optometry}} \bibinfo{volume}{95},
  \bibinfo{number}{12} (\bibinfo{date}{December} \bibinfo{year}{2018}),
  \bibinfo{pages}{1135—1141}.
\newblock
\showISSN{1040-5488}
\urldef\tempurl%
\url{https://doi.org/10.1097/opx.0000000000001308}
\showDOI{\tempurl}


\bibitem[Jesus et~al\mbox{.}(2016)]%
        {spot_clinical}
\bibfield{author}{\bibinfo{person}{Daniela Jesus}, \bibinfo{person}{Flávio
  Villela}, \bibinfo{person}{Luis Orlandin}, \bibinfo{person}{Fernando Eiji},
  \bibinfo{person}{Daniel Dantas}, {and} \bibinfo{person}{Milton Alves}.}
  \bibinfo{year}{2016}\natexlab{}.
\newblock \showarticletitle{Comparison between refraction measured by Spot
  Vision ScreeningTM and subjective clinical refractometry}.
\newblock \bibinfo{journal}{\emph{Clinics}}  \bibinfo{volume}{70}
  (\bibinfo{date}{02} \bibinfo{year}{2016}), \bibinfo{pages}{69--72}.
\newblock
\urldef\tempurl%
\url{https://doi.org/10.6061/clinics/2016(02)03}
\showDOI{\tempurl}


\bibitem[Jorge et~al\mbox{.}(2005)]%
        {retino_vs_ar_1}
\bibfield{author}{\bibinfo{person}{Jorge Jorge}, \bibinfo{person}{António
  Queirós}, \bibinfo{person}{José~B Almeida}, {and} \bibinfo{person}{Manuel~A
  Parafita}.} \bibinfo{year}{2005}\natexlab{}.
\newblock \showarticletitle{Retinoscopy/autorefraction: which is the best
  starting point for a noncycloplegic refraction?}
\newblock \bibinfo{journal}{\emph{Optometry and vision science : official
  publication of the American Academy of Optometry}} \bibinfo{volume}{82},
  \bibinfo{number}{1} (\bibinfo{date}{January} \bibinfo{year}{2005}),
  \bibinfo{pages}{64—68}.
\newblock
\showISSN{1040-5488}
\urldef\tempurl%
\url{http://europepmc.org/abstract/MED/15630406}
\showURL{%
\tempurl}


\bibitem[Kullback and Leibler(1951)]%
        {KLD}
\bibfield{author}{\bibinfo{person}{S. Kullback} {and} \bibinfo{person}{R.~A.
  Leibler}.} \bibinfo{year}{1951}\natexlab{}.
\newblock \showarticletitle{On Information and Sufficiency}.
\newblock \bibinfo{journal}{\emph{The Annals of Mathematical Statistics}}
  \bibinfo{volume}{22}, \bibinfo{number}{1} (\bibinfo{year}{1951}),
  \bibinfo{pages}{79--86}.
\newblock
\showISSN{00034851}


\bibitem[Lawless et~al\mbox{.}(2007)]%
        {retino_kerato_2}
\bibfield{author}{\bibinfo{person}{Michael Lawless}, \bibinfo{person}{FRACS
  FRACO}, \bibinfo{person}{Anthony DCLP}, {and} \bibinfo{person}{Mark FRACS}.}
  \bibinfo{year}{2007}\natexlab{}.
\newblock \showarticletitle{Keratoconus: Diagnosis and management}.
\newblock \bibinfo{journal}{\emph{Australian and New Zealand Journal of
  Ophthalmology}}  \bibinfo{volume}{17} (\bibinfo{date}{11}
  \bibinfo{year}{2007}), \bibinfo{pages}{33 -- 60}.
\newblock
\urldef\tempurl%
\url{https://doi.org/10.1111/j.1442-9071.1989.tb00487.x}
\showDOI{\tempurl}


\bibitem[Lee(2015)]%
        {aao-retinoscopy}
\bibfield{author}{\bibinfo{person}{Olivia~L Lee}.}
  \bibinfo{year}{2015}\natexlab{}.
\newblock \bibinfo{booktitle}{\emph{Retinoscopy 101}}.
\newblock
\urldef\tempurl%
\url{https://www.aao.org/young-ophthalmologists/yo-info/article/retinoscopy-101}
\showURL{%
Retrieved February 1,2022 from \tempurl}


\bibitem[Leone~(nee Major) et~al\mbox{.}(2010)]%
        {subjective_hyperopia_challenge}
\bibfield{author}{\bibinfo{person}{Jody Leone~(nee Major)},
  \bibinfo{person}{Paul Mitchell}, \bibinfo{person}{Ian Morgan},
  \bibinfo{person}{Annette Kifley}, {and} \bibinfo{person}{Kathryn Rose}.}
  \bibinfo{year}{2010}\natexlab{}.
\newblock \showarticletitle{Use of Visual Acuity to Screen for Significant
  Refractive Errors in Adolescents Is It Reliable?}
\newblock \bibinfo{journal}{\emph{Archives of ophthalmology}}
  \bibinfo{volume}{128} (\bibinfo{date}{07} \bibinfo{year}{2010}),
  \bibinfo{pages}{894--9}.
\newblock
\urldef\tempurl%
\url{https://doi.org/10.1001/archophthalmol.2010.134}
\showDOI{\tempurl}


\bibitem[Lukežic et~al\mbox{.}(2017)]%
        {CSRT_tracker}
\bibfield{author}{\bibinfo{person}{Alan Lukežic}, \bibinfo{person}{Tomáš
  Vojír}, \bibinfo{person}{Luka~Cehovin Zajc}, \bibinfo{person}{Jirí Matas},
  {and} \bibinfo{person}{Matej Kristan}.} \bibinfo{year}{2017}\natexlab{}.
\newblock \showarticletitle{Discriminative Correlation Filter with Channel and
  Spatial Reliability}. In \bibinfo{booktitle}{\emph{2017 IEEE Conference on
  Computer Vision and Pattern Recognition (CVPR)}}.
  \bibinfo{pages}{4847--4856}.
\newblock
\urldef\tempurl%
\url{https://doi.org/10.1109/CVPR.2017.515}
\showDOI{\tempurl}


\bibitem[Mariakakis et~al\mbox{.}(2017)]%
        {BiliScreen}
\bibfield{author}{\bibinfo{person}{Alex Mariakakis}, \bibinfo{person}{Megan~A.
  Banks}, \bibinfo{person}{Lauren Phillipi}, \bibinfo{person}{Lei Yu},
  \bibinfo{person}{James Taylor}, {and} \bibinfo{person}{Shwetak~N. Patel}.}
  \bibinfo{year}{2017}\natexlab{}.
\newblock \showarticletitle{BiliScreen: Smartphone-Based Scleral Jaundice
  Monitoring for Liver and Pancreatic Disorders}.
\newblock \bibinfo{journal}{\emph{Proc. ACM Interact. Mob. Wearable Ubiquitous
  Technol.}} \bibinfo{volume}{1}, \bibinfo{number}{2}, Article
  \bibinfo{articleno}{20} (\bibinfo{date}{jun} \bibinfo{year}{2017}),
  \bibinfo{numpages}{26}~pages.
\newblock
\urldef\tempurl%
\url{https://doi.org/10.1145/3090085}
\showDOI{\tempurl}


\bibitem[McCullough et~al\mbox{.}(2017)]%
        {PMID:28030881}
\bibfield{author}{\bibinfo{person}{Sara~J McCullough}, \bibinfo{person}{Lesley
  Doyle}, {and} \bibinfo{person}{Kathryn~J Saunders}.}
  \bibinfo{year}{2017}\natexlab{}.
\newblock \showarticletitle{Intra- and inter- examiner repeatability of
  cycloplegic retinoscopy among young children}.
\newblock \bibinfo{journal}{\emph{Ophthalmic \& physiological optics : the
  journal of the British College of Ophthalmic Opticians (Optometrists)}}
  \bibinfo{volume}{37}, \bibinfo{number}{1} (\bibinfo{date}{January}
  \bibinfo{year}{2017}), \bibinfo{pages}{16—23}.
\newblock
\showISSN{0275-5408}
\urldef\tempurl%
\url{https://doi.org/10.1111/opo.12341}
\showDOI{\tempurl}


\bibitem[Murali et~al\mbox{.}(2021)]%
        {kanna}
\bibfield{author}{\bibinfo{person}{Kaushik Murali}, \bibinfo{person}{Viswesh
  Krishna}, \bibinfo{person}{Vrishab Krishna}, \bibinfo{person}{B. Kumari},
  \bibinfo{person}{Sowmya~Raveendra Murthy}, \bibinfo{person}{Vidhya C}, {and}
  \bibinfo{person}{Payal Shah}.} \bibinfo{year}{2021}\natexlab{}.
\newblock \showarticletitle{Effectiveness of Kanna photoscreener in detecting
  amblyopia risk factors}.
\newblock \bibinfo{journal}{\emph{Indian Journal of Ophthalmology}}
  \bibinfo{volume}{69}, \bibinfo{number}{8} (\bibinfo{year}{2021}).
\newblock
\showISSN{0301-4738}
\urldef\tempurl%
\url{https://journals.lww.com/ijo/Fulltext/2021/08000/Effectiveness_of_Kanna_photoscreener_in_detecting.19.aspx}
\showURL{%
\tempurl}


\bibitem[Naidoo et~al\mbox{.}(2016)]%
        {Naidoo2016-hc}
\bibfield{author}{\bibinfo{person}{Kovin~S Naidoo}, \bibinfo{person}{Janet
  Leasher}, \bibinfo{person}{Rupert~R Bourne}, \bibinfo{person}{Seth~R
  Flaxman}, \bibinfo{person}{Jost~B Jonas}, \bibinfo{person}{Jill Keeffe},
  \bibinfo{person}{Hans Limburg}, \bibinfo{person}{Konrad Pesudovs},
  \bibinfo{person}{Holly Price}, \bibinfo{person}{Richard~A White},
  \bibinfo{person}{Tien~Y Wong}, \bibinfo{person}{Hugh~R Taylor},
  \bibinfo{person}{Serge Resnikoff}, {and} \bibinfo{person}{{Vision Loss Expert
  Group of the Global Burden of Disease Study}}.}
  \bibinfo{year}{2016}\natexlab{}.
\newblock \showarticletitle{Global Vision Impairment and Blindness Due to
  Uncorrected Refractive Error, 1990-2010}.
\newblock \bibinfo{journal}{\emph{Optom Vis Sci}} \bibinfo{volume}{93},
  \bibinfo{number}{3} (\bibinfo{date}{March} \bibinfo{year}{2016}),
  \bibinfo{pages}{227--234}.
\newblock


\bibitem[Organization(2019)]%
        {WHO_vision_report_2019}
\bibfield{author}{\bibinfo{person}{World~Health Organization}.}
  \bibinfo{year}{2019}\natexlab{}.
\newblock \bibinfo{booktitle}{\emph{World report on vision}}.
\newblock \bibinfo{publisher}{World Health Organization}. 160 p. pages.
\newblock


\bibitem[Otsu(1979)]%
        {OTSU}
\bibfield{author}{\bibinfo{person}{Nobuyuki Otsu}.}
  \bibinfo{year}{1979}\natexlab{}.
\newblock \showarticletitle{A threshold selection method from gray level
  histograms}.
\newblock \bibinfo{journal}{\emph{IEEE Transactions on Systems, Man, and
  Cybernetics}}  \bibinfo{volume}{9} (\bibinfo{year}{1979}),
  \bibinfo{pages}{62--66}.
\newblock


\bibitem[Owen(2007)]%
        {Huber}
\bibfield{author}{\bibinfo{person}{Art Owen}.} \bibinfo{year}{2007}\natexlab{}.
\newblock \showarticletitle{A robust hybrid of lasso and ridge regression}.
\newblock \bibinfo{journal}{\emph{Contemp. Math.}}  \bibinfo{volume}{443}
  (\bibinfo{date}{01} \bibinfo{year}{2007}).
\newblock
\showISBNx{9780821841952}
\urldef\tempurl%
\url{https://doi.org/10.1090/conm/443/08555}
\showDOI{\tempurl}


\bibitem[Pamplona et~al\mbox{.}(2010)]%
        {Netra}
\bibfield{author}{\bibinfo{person}{Vitor~F. Pamplona}, \bibinfo{person}{Ankit
  Mohan}, \bibinfo{person}{Manuel~M. Oliveira}, {and} \bibinfo{person}{Ramesh
  Raskar}.} \bibinfo{year}{2010}\natexlab{}.
\newblock \showarticletitle{NETRA: Interactive Display for Estimating
  Refractive Errors and Focal Range}.
\newblock \bibinfo{journal}{\emph{ACM Trans. Graph.}} \bibinfo{volume}{29},
  \bibinfo{number}{4}, Article \bibinfo{articleno}{77} (\bibinfo{date}{jul}
  \bibinfo{year}{2010}), \bibinfo{numpages}{8}~pages.
\newblock
\showISSN{0730-0301}
\urldef\tempurl%
\url{https://doi.org/10.1145/1778765.1778814}
\showDOI{\tempurl}


\bibitem[Park et~al\mbox{.}(2021)]%
        {rdtscan}
\bibfield{author}{\bibinfo{person}{Chunjong Park}, \bibinfo{person}{Hung Ngo},
  \bibinfo{person}{Libby~Rose Lavitt}, \bibinfo{person}{Vincent Karuri},
  \bibinfo{person}{Shiven Bhatt}, \bibinfo{person}{Peter Lubell-Doughtie},
  \bibinfo{person}{Anuraj~H. Shankar}, \bibinfo{person}{Leonard Ndwiga},
  \bibinfo{person}{Victor Osoti}, \bibinfo{person}{Juliana~K. Wambua},
  \bibinfo{person}{Philip Bejon}, \bibinfo{person}{Lynette~Isabella
  Ochola-Oyier}, \bibinfo{person}{Monique Chilver}, \bibinfo{person}{Nigel
  Stocks}, \bibinfo{person}{Victoria Lyon}, \bibinfo{person}{Barry~R. Lutz},
  \bibinfo{person}{Matthew Thompson}, \bibinfo{person}{Alex Mariakakis}, {and}
  \bibinfo{person}{Shwetak Patel}.} \bibinfo{year}{2021}\natexlab{}.
\newblock \showarticletitle{The Design and Evaluation of a Mobile System for
  Rapid Diagnostic Test Interpretation}.
\newblock \bibinfo{journal}{\emph{Proc. ACM Interact. Mob. Wearable Ubiquitous
  Technol.}} \bibinfo{volume}{5}, \bibinfo{number}{1}, Article
  \bibinfo{articleno}{29} (\bibinfo{date}{mar} \bibinfo{year}{2021}),
  \bibinfo{numpages}{26}~pages.
\newblock
\urldef\tempurl%
\url{https://doi.org/10.1145/3448106}
\showDOI{\tempurl}


\bibitem[Pascolini and Mariotti(2012)]%
        {Pascolini614}
\bibfield{author}{\bibinfo{person}{Donatella Pascolini} {and}
  \bibinfo{person}{Silvio~Paolo Mariotti}.} \bibinfo{year}{2012}\natexlab{}.
\newblock \showarticletitle{Global estimates of visual impairment: 2010}.
\newblock \bibinfo{journal}{\emph{British Journal of Ophthalmology}}
  \bibinfo{volume}{96}, \bibinfo{number}{5} (\bibinfo{year}{2012}),
  \bibinfo{pages}{614--618}.
\newblock
\showISSN{0007-1161}
\urldef\tempurl%
\url{https://doi.org/10.1136/bjophthalmol-2011-300539}
\showDOI{\tempurl}


\bibitem[Payerols et~al\mbox{.}(2016)]%
        {plusoptix_clinical}
\bibfield{author}{\bibinfo{person}{Arnaud Payerols}, \bibinfo{person}{Claudie
  Eliaou}, \bibinfo{person}{Véronique Trezeguet}, \bibinfo{person}{Max
  Villain}, {and} \bibinfo{person}{Vincent Daien}.}
  \bibinfo{year}{2016}\natexlab{}.
\newblock \showarticletitle{Accuracy of PlusOptix A09 distance refraction in
  pediatric myopia and hyperopia}.
\newblock \bibinfo{journal}{\emph{BMC ophthalmology}}  \bibinfo{volume}{16}
  (\bibinfo{date}{June} \bibinfo{year}{2016}), \bibinfo{pages}{72}.
\newblock
\showISSN{1471-2415}
\urldef\tempurl%
\url{https://doi.org/10.1186/s12886-016-0247-8}
\showDOI{\tempurl}


\bibitem[Rahi et~al\mbox{.}(2008)]%
        {uk-reference}
\bibfield{author}{\bibinfo{person}{J~S Rahi}, \bibinfo{person}{C~S Peckham},
  {and} \bibinfo{person}{P~M Cumberland}.} \bibinfo{year}{2008}\natexlab{}.
\newblock \showarticletitle{Visual impairment due to undiagnosed refractive
  error in working age adults in Britain}.
\newblock \bibinfo{journal}{\emph{Br J Ophthalmol}} \bibinfo{volume}{92},
  \bibinfo{number}{9} (\bibinfo{date}{Sept.} \bibinfo{year}{2008}),
  \bibinfo{pages}{1190--1194}.
\newblock


\bibitem[Roorda et~al\mbox{.}(1997)]%
        {eccentric_photorefraction_slope}
\bibfield{author}{\bibinfo{person}{Austin Roorda}, \bibinfo{person}{Melanie
  Campbell}, {and} \bibinfo{person}{William Bobier}.}
  \bibinfo{year}{1997}\natexlab{}.
\newblock \showarticletitle{Slope-based eccentric photorefraction: Theoretical
  analysis of different light source configurations and effects of ocular
  aberrations}.
\newblock \bibinfo{journal}{\emph{Journal of the Optical Society of America. A,
  Optics, image science, and vision}}  \bibinfo{volume}{14} (\bibinfo{date}{11}
  \bibinfo{year}{1997}), \bibinfo{pages}{2547--56}.
\newblock
\urldef\tempurl%
\url{https://doi.org/10.1364/JOSAA.14.002547}
\showDOI{\tempurl}


\bibitem[Safir et~al\mbox{.}(1970)]%
        {retinoscopy_precision}
\bibfield{author}{\bibinfo{person}{Aran Safir}, \bibinfo{person}{Lyon Hyams},
  \bibinfo{person}{John Philpot}, {and} \bibinfo{person}{Louis~S. Jagerman}.}
  \bibinfo{year}{1970}\natexlab{}.
\newblock \showarticletitle{{Studies in Refraction: I. The Precision of
  Retinoscopy}}.
\newblock \bibinfo{journal}{\emph{Archives of Ophthalmology}}
  \bibinfo{volume}{84}, \bibinfo{number}{1} (\bibinfo{date}{07}
  \bibinfo{year}{1970}), \bibinfo{pages}{49--61}.
\newblock
\showISSN{0003-9950}
\urldef\tempurl%
\url{https://doi.org/10.1001/archopht.1970.00990040051013}
\showDOI{\tempurl}


\bibitem[Sainani(2013)]%
        {coopeartive_challenges_AR}
\bibfield{author}{\bibinfo{person}{Ashwin Sainani}.}
  \bibinfo{year}{2013}\natexlab{}.
\newblock \showarticletitle{Special considerations for prescription of glasses
  in children}.
\newblock \bibinfo{journal}{\emph{Journal of Clinical Ophthalmology and
  Research}}  \bibinfo{volume}{1} (\bibinfo{date}{01} \bibinfo{year}{2013}),
  \bibinfo{pages}{169}.
\newblock
\urldef\tempurl%
\url{https://doi.org/10.4103/2320-3897.116861}
\showDOI{\tempurl}


\bibitem[Sravani et~al\mbox{.}(2015)]%
        {eccentric_ethnicity_challenge}
\bibfield{author}{\bibinfo{person}{Geetha Sravani}, \bibinfo{person}{Vinay
  Nilagiri}, {and} \bibinfo{person}{Shrikant Bharadwaj}.}
  \bibinfo{year}{2015}\natexlab{}.
\newblock \showarticletitle{Photorefraction estimates of refractive power
  varies with the ethnic origin of human eyes}.
\newblock \bibinfo{journal}{\emph{Scientific reports}}  \bibinfo{volume}{5}
  (\bibinfo{date}{01} \bibinfo{year}{2015}), \bibinfo{pages}{7976}.
\newblock
\urldef\tempurl%
\url{https://doi.org/10.1038/srep07976}
\showDOI{\tempurl}


\bibitem[Tielsch et~al\mbox{.}(1990)]%
        {us-referror}
\bibfield{author}{\bibinfo{person}{J~M Tielsch}, \bibinfo{person}{A Sommer},
  \bibinfo{person}{K Witt}, \bibinfo{person}{J Katz}, {and}
  \bibinfo{person}{R~M Royall}.} \bibinfo{year}{1990}\natexlab{}.
\newblock \showarticletitle{Blindness and visual impairment in an American
  urban population. The Baltimore Eye Survey}.
\newblock \bibinfo{journal}{\emph{Arch Ophthalmol}} \bibinfo{volume}{108},
  \bibinfo{number}{2} (\bibinfo{date}{Feb.} \bibinfo{year}{1990}),
  \bibinfo{pages}{286--290}.
\newblock


\bibitem[Ugurbas et~al\mbox{.}(2019)]%
        {plusoptix_eval}
\bibfield{author}{\bibinfo{person}{Silay Ugurbas}, \bibinfo{person}{Numan
  Küçük}, \bibinfo{person}{Irem Isik}, \bibinfo{person}{Atilla Alpay},
  \bibinfo{person}{Cagatay Buyukuysal}, {and} \bibinfo{person}{Suat
  Uğurbaş}.} \bibinfo{year}{2019}\natexlab{}.
\newblock \showarticletitle{Objective vision screening using PlusoptiX for
  children aged 3–11 years in rural Turkey}.
\newblock \bibinfo{journal}{\emph{BMC Ophthalmology}}  \bibinfo{volume}{19}
  (\bibinfo{date}{03} \bibinfo{year}{2019}).
\newblock
\urldef\tempurl%
\url{https://doi.org/10.1186/s12886-019-1080-7}
\showDOI{\tempurl}


\bibitem[Wang et~al\mbox{.}(2018)]%
        {seismo}
\bibfield{author}{\bibinfo{person}{Edward~Jay Wang}, \bibinfo{person}{Junyi
  Zhu}, \bibinfo{person}{Mohit Jain}, \bibinfo{person}{Tien-Jui Lee},
  \bibinfo{person}{Elliot Saba}, \bibinfo{person}{Lama Nachman}, {and}
  \bibinfo{person}{Shwetak~N. Patel}.} \bibinfo{year}{2018}\natexlab{}.
\newblock \bibinfo{booktitle}{\emph{Seismo: Blood Pressure Monitoring Using
  Built-in Smartphone Accelerometer and Camera}}.
\newblock \bibinfo{publisher}{Association for Computing Machinery},
  \bibinfo{address}{New York, NY, USA}, \bibinfo{pages}{1–9}.
\newblock
\showISBNx{9781450356206}
\urldef\tempurl%
\url{https://doi.org/10.1145/3173574.3173999}
\showURL{%
\tempurl}


\bibitem[Wesemann et~al\mbox{.}(1992)]%
        {eccentric_photorefraction_theory}
\bibfield{author}{\bibinfo{person}{Wolfgang Wesemann}, \bibinfo{person}{Anthony
  Norcia}, {and} \bibinfo{person}{Dale Allen}.}
  \bibinfo{year}{1992}\natexlab{}.
\newblock \showarticletitle{Theory of eccentric photorefraction
  (photoretinoscopy): astigmatic eyes}.
\newblock \bibinfo{journal}{\emph{Journal of the Optical Society of America. A,
  Optics and image science}}  \bibinfo{volume}{8} (\bibinfo{date}{01}
  \bibinfo{year}{1992}), \bibinfo{pages}{2038--47}.
\newblock
\urldef\tempurl%
\url{https://doi.org/10.1364/JOSAA.8.002038}
\showDOI{\tempurl}


\bibitem[West and Williams(2016)]%
        {amblyopia_age}
\bibfield{author}{\bibinfo{person}{Stephanie West} {and} \bibinfo{person}{Cathy
  Williams}.} \bibinfo{year}{2016}\natexlab{}.
\newblock \showarticletitle{Amblyopia in children (aged 7 years or less)}.
\newblock \bibinfo{journal}{\emph{BMJ clinical evidence}}
  \bibinfo{volume}{2016} (\bibinfo{date}{01} \bibinfo{year}{2016}).
\newblock


\bibitem[Yang et~al\mbox{.}(2020)]%
        {smartphone_1_photorefraction}
\bibfield{author}{\bibinfo{person}{Zhongqi Yang}, \bibinfo{person}{Eugene Fu},
  \bibinfo{person}{Grace Ngai}, \bibinfo{person}{Hong Leong},
  \bibinfo{person}{Chi~Wai Do}, {and} \bibinfo{person}{Lily Chan}.}
  \bibinfo{year}{2020}\natexlab{}.
\newblock \showarticletitle{Screening for refractive error with low-quality
  smartphone images}. \bibinfo{pages}{119--128}.
\newblock
\urldef\tempurl%
\url{https://doi.org/10.1145/3428690.3429175}
\showDOI{\tempurl}


\bibitem[Zhou(2015)]%
        {sv_one_AR}
\bibfield{author}{\bibinfo{person}{Yaopeng Zhou}.}
  \bibinfo{year}{2015}\natexlab{}.
\newblock \showarticletitle{Portable wavefront aberrometer}.
\newblock  \bibinfo{number}{9066683} (\bibinfo{date}{06} \bibinfo{year}{2015}).
\newblock
\urldef\tempurl%
\url{https://www.freepatentsonline.com/9066683.html}
\showURL{%
\tempurl}


\bibitem[Zitelli et~al\mbox{.}(2012)]%
        {Optham_book}
\bibfield{author}{\bibinfo{person}{Basil~J. Zitelli}, \bibinfo{person}{Sara~C.
  McIntire}, {and} \bibinfo{person}{Andrew~J. Nowalk}.}
  \bibinfo{year}{2012}\natexlab{}.
\newblock \bibinfo{booktitle}{\emph{Zitelli and Davis' atlas of pediatric
  physical diagnosis}}.
\newblock \bibinfo{publisher}{Saunders/Elsevier}.
\newblock
\showISBNx{9780323393034}


\bibitem[Zuiderveld(1994)]%
        {CLAHE}
\bibfield{author}{\bibinfo{person}{Karel~J. Zuiderveld}.}
  \bibinfo{year}{1994}\natexlab{}.
\newblock \showarticletitle{Contrast Limited Adaptive Histogram Equalization}.
  In \bibinfo{booktitle}{\emph{Graphics Gems}}.
\newblock


\end{thebibliography}
